\documentclass[preprint,aps,showpacs]{revtex4}
\usepackage{mathrsfs}
\usepackage{amsmath}
\usepackage{amssymb}
\usepackage{epsfig}
\usepackage{graphicx}
\usepackage{booktabs}

\begin{document}
\renewcommand{\arraystretch}{0.5}
\newcommand{\beq}{\begin{eqnarray}}
\newcommand{\eeq}{\end{eqnarray}}
\newcommand{\non}{\nonumber\\ }

\newcommand{\acp}{ {\cal A}_{CP} }
\newcommand{\psl}{ p \hspace{-1.8truemm}/ }
\newcommand{\nsl}{ n \hspace{-2.2truemm}/ }
\newcommand{\vsl}{ v \hspace{-2.2truemm}/ }
\newcommand{\epsl}{\epsilon \hspace{-1.8truemm}/\,  }

\def \cpl{ Chin. Phys. Lett.  }
\def \ctp{ Commun. Theor. Phys.  }
\def \epjc{ Eur. Phys. J. C }
\def \jpg{  J. Phys. G }
\def \npb{  Nucl. Phys. B }
\def \plb{  Phys. Lett. B }
\def \prd{  Phys. Rev. D }
\def \prl{  Phys. Rev. Lett.  }
\def \zpc{  Z. Phys. C }
\def \jhep{ J. High Energy Phys.  }

\title{ The two-body   $B_c\rightarrow D_{(s)}^{(*)}P$, $D_{(s)}^{(*)}V$ decays  in the
Perturbative QCD Approach}
\author{Zhou Rui }
\author{ Zhi-Tian Zou}
\author{Cai-Dian L\"{u}}\email{lucd@ihep.ac.cn}
\affiliation{Institute  of  High  Energy  Physics  and  Theoretical  Physics Center for Science Facilities,
Chinese Academy of Sciences, Beijing 100049, People's Republic of China }

\date{\today}
\begin{abstract}
We make a systematic  investigation on the two-body  nonleptonic
decays $B_c\rightarrow D_{(s)}^{(*)}P$, $D_{(s)}^{(*)}V$, by
employing the perturbative QCD   approach based on $k_T$
factorization, where P and V denote any light pseudoscalar meson and
vector meson, respectively. We  predict the branching ratios and
direct CP-asymmetries of these $B_c$ decays
 and also   the  transverse
polarization fractions of  $B_c \to D_{(s)}^* V$ decays. It is found
that the non-factorizable emission diagrams and annihilation type
diagrams have remarkable effects on the physical observables in many
channels, especially the color-suppressed and annihilation-dominant
decay modes. A possible large direct CP-violation is predicted in
some channels; and a  large transverse polarization contribution
that can reach $50\% \sim70\%$ is predicted in some of the
$B_c\rightarrow D_{(s)}^*V$ decays.
\end{abstract}



\maketitle

\section{Introduction}
The $B_c$ meson is the only quark-antiquark bound system
$(\bar{b}c)$ composed of both heavy quarks  with different flavors,
and are thus flavor asymmetric.  It can decay only via weak
interaction, since the two flavor asymmetric quarks (b and c) can
not annihilate into gluons or photon  via strong interaction or
electromagnetic interaction.  Because  each of the two heavy quarks
can decay individually, and they can also annihilate through weak
interaction, $B_c$ meson has rich decay channels and provides a
very good place to study nonleptonic weak decays of heavy mesons, to
test the standard model and to search for any new physics signals \cite{iiba}.

Since the current running LHC collider will produce much more $B_c$
mesons than ever before, a lot of theoretical studies of the
nonleptonic $B_c$ weak decays have been performed using different
approaches. For example, the spectator-model \cite{prd391342}, the
light-front quark model (LFQM) \cite{09095028,prd80054016}, the relativistic
constituent quark model (RCQM) \cite{prd73054024}, the QCD
factorization approach (QCDF) \cite {prd77114004}, the Perturbative
QCD approach (pQCD) \cite{prd81014022,epjc45711,epjc60107,epjc63435},
and so on.
Among the numerous decay channels, there is one category with only
one charmed meson in the final states. They are rare decays, but
with possible large direct CP asymmetry, since there are both
penguin and tree diagrams involved. These decays have ever been
studied
  in Ref. \cite{09095028} using the naive factorization approach. But they   consider
only the contribution of current-current operators at the tree level, and
thus no direct CP asymmetry is predicted. They also have difficulty
to predict those     pure penguin type or    annihilation  dominant  type decays, such as
$B_c\rightarrow D^+\phi$, $D_s^+\bar{K}^0$, $D_s^+\phi$. Ref.
\cite{prd73054024} discussed some semileptonic and nonleptonic $B_c$
weak decays and CP-violating asymmetries by using RCQM model based
on the Bethe-Salpeter formalism. They  do not include the
contributions of  annihilation type diagrams, either. Since  the
annihilation type contributions are found to be important in the $B$
meson non-leptonic decays \cite{chenghaiyang} and also significant
in the $B_c$ decays  \cite{epjc5705}, one needs further study these
channels carefully.

 In this paper, we calculate all the processes of a $B_c$ meson
decays to one $D^{(*)}_{(s)}$ meson and one light pseudoscalar meson
(P) or vector meson (V) in pQCD approach. It is well-known that $B_c$ meson is
a nonrelativistic heavy quarkonium system.  Thus the two quarks in the $B_c$ meson are both at rest and non-relativistic.  Since the charm quark in the final state $D$ meson is almost at collinear state,  a hard gluon is needed to transfer large momentum to the spectator charm quark. 
In the leading order of $m_c/m_{B_c} \sim 0.2$ expansion,     the   factorization theorem  is
applicable to the $B_c$ system  similar to the situation of  B meson
\cite{prd84074033}.  Utilizing  the $k_T$
factorization instead of collinear factorization, this approach is
free of endpoint singularity. Thus the  diagrams including
factorizable, nonfactorizable and annihilation type, are all
calculable. It has
  been tested  in the study of charmless $B$ meson decays  successfully \cite{prl744388}, especially for the direct CP
asymmetries \cite{0505020}.   For the charmed  decays of $B$ meson,
 it is also demonstrated to be
  applicable in the leading order of the $m_D/m_B$ expansion \cite{08031073,09081856,0512347,0305335,0112127,prd67054028}.

Our paper is organized as follows: We review the pQCD factorization
approach and then   perform the perturbative calculations for these
considered decay channels   in Sec.\ref{sec:f-work}.  The numerical
results and discussions on the observables are given in
Sec.\ref{sec:numerical}. The final section is devoted to our
conclusions. Some details related functions and the decay amplitudes
are given in the Appendix.

\section{ Theoretical framework}\label{sec:f-work}

For the charmed $B_c$ decays we considered, the weak effective
Hamiltonian $\mathcal {H}_{eff}$ for $b\rightarrow q'(q'=d,s)$
transition can be written as \cite{rmp681125}
\begin{eqnarray}\label{eq:heff}
\mathcal
{H}_{eff}=\frac{G_F}{\sqrt{2}}\{\sum_{q=u,c}\xi_{q}[(C_1(\mu)O_1^q(\mu)
+C_2(\mu)O_2^q(\mu))+\sum_{i=3}^{10}C_i(\mu)O_i(\mu)]\},
\end{eqnarray}
with the Cabbibo-Kobayashi-Maskawa (CKM) matrix element
$\xi_{q}=V_{qq'}V^*_{qb}$. $O_i(\mu)$ and $C_i(\mu)$
 are the effective four quark operators and their QCD corrected Wilson coefficients, respectively.
 Their  expressions  can be found easily for example
in Ref. \cite{rmp681125}.

With these quark level weak operators, the hardest work is left for
the matrix element calculation between hadronic states $\langle D M
| H_{eff}|B_c\rangle$. Since both perturbative and non-perturbative
QCD are involved, the factorization theorem is required to make the
calculation meaningful. The perturbative QCD approach
\cite{prl744388} is one of the methods to deal with hadronic B
decays based on $k_T$ factorization. 
 At zero  recoil of D meson in the semi-leptonic $B_c$ decay, both c and b quark can be described by heavy quark effective theory. However, when the D meson is at maximum recoil, which is the case of two body non-leptonic $B_c$ decay,  the final state
mesons at the rest frame of $B_c$ meson are collinear, so as to the
constituent quarks (c and other light quarks) inside. Since  the spectator $c$ quark in the
$B_c$ meson is almost at rest,
a hard gluon is then needed to
transform it into a collinear object in the final state meson. This
makes the perturbative calculations into a 6-quark interaction. In
this collinear factorization calculation, endpoint singularity
usually appears in some of the diagrams. The QCD factorization
approach \cite{bbns} just parameterize those diagrams with
singularity as free parameters; while in the so-called
soft-collinear effective theory \cite{scet}, people separate these
incalculable part to an unknown matrix element. In our pQCD
approach, we studied these singularity and found that they arise
from the endpoint where longitudinal momentum is small. Therefore,
the transverse momentum of quarks is no longer negligible. If one
pick back the transverse momentum, the result is finite.


Because the intrinsic transverse momentum of quarks is smaller than
the $b$ quark mass scale, therefore we have one more scale than the
usual collinear factorization. Additional double logarithms appear
at the perturbative QCD calculations. These large logarithms will
spoil the perturbation expansion, thus a resummation is required.
This has been done to give the so called Sudakov form factors
\cite{npb193381}. The single logarithm between the W boson mass
scale and the factorization scale t in pQCD approach has been
absorbed into the Wilson coefficients of four quark operators. The
decay amplitude is then factorized into the convolution of the hard
subamplitude, the Wilson coefficient and the Sudakov factor with the
meson wave functions, all of which are well-defined and gauge
invariant. Therefore, the three-scale factorization formula for
exclusive nonleptonic $B$ meson decays is then written as
\begin{eqnarray}\label{eq:factorization}
C(t)\otimes H(x,t)\otimes
\Phi(x)\otimes\exp\left[-s(P,b)-2\int^t_{1/b}\frac{d\mu}{\mu}\gamma_q(\alpha_s(\mu))\right],
\end{eqnarray}
where $C(t)$ are the corresponding Wilson coefficients. The Sudakov
evolution $\exp[-s(P, b)]$ \cite{npb193381} are from the resummation
of double logarithms $\ln^2 (Pb)$, with P denoting the dominant
light-cone component of meson momentum. $\gamma_q=-\alpha_s/\pi$ is
the quark anomalous dimension in axial gauge. All non-perturbative
components are organized in the form of hadron wave functions
$\Phi(x)$, which can be extracted from experimental data or other
non-perturbative method.  Since non-perturbative dynamics has been
factored out, one can evaluate all possible Feynman diagrams for the
six-quark amplitude straightforwardly, which include both traditional
factorizable and so-called ``non-factorizable'' contributions. Factorizable and
non-factorizable annihilation type diagrams are also calculable
without endpoint singularity.

The meson wave function,
 which describes hadronization of the quark and anti-quark
inside the meson, is independent of the specific processes. Using
the wave functions determined from other well measured processes,
one can make quantitative predictions here. For the light
pseudoscalar meson, its wave function can be defined
as \cite{zpc48239}
\begin{eqnarray}\label{eq:piwave}
\Phi(P,x,\xi)=\frac{i}{2N_c}\gamma_5[\rlap{/}{P}\phi_P^A(x)+m_0\phi_P^P(x)+\xi
m_0(\rlap{/}{n}\rlap{/}{v}-1)\phi^T_P(x)],
\end{eqnarray}
where $ P$  is the momentum of the light meson, and $x$ is the
momentum fraction of the quark (or anti-quark) inside the meson. When the momentum fraction of the quark (anti-quark)
is set to be $x$, the parameter $\xi$ should be chosen as $+1 (-1)$.
The distribution amplitudes $\phi^A_P(x),\phi_P^P(x)$ and $\phi_P^T(x)$
are given in Appendix \ref{sec:c}.

For the light vector mesons, both longitudes (L) and transverse (T)
polarizations are involved. Their wave functions are  written
as \cite{prd81014022}
\begin{eqnarray}
\Phi_V^L(x)&=&\frac{1}{\sqrt{2N_c}}\{
M_V\rlap{/}{\epsilon}_V^{*L}\phi_V(x)+
\rlap{/}{\epsilon}_V^{*L}\rlap{/}{P}\phi_V^t(x)+M_V\phi_V^s(x)\}_{\alpha\beta},
\nonumber\\\Phi_V^T(x)&=&\frac{1}{\sqrt{2N_c}}\{
M_V\rlap{/}{\epsilon}_V^{*T}\phi^{\nu}_V(x)+
\rlap{/}{\epsilon}_V^{*T}\rlap{/}{P}\phi_V^T(x)+
iM_V\epsilon_{\mu\nu\rho\sigma}\gamma_5
\epsilon_T^{*\nu}n^{\rho}v^{\sigma}\phi_V^a(x)\}_{\alpha\beta},
\end{eqnarray}
where $\epsilon_V^{L(T)}$ denotes the longitudinal (transverse)
polarization vector. And convention $\epsilon^{0123}=1$ is adopted
for the Levi-Civita tensor. The distributions amplitudes are also
presented in Appendix \ref{sec:c}.

 Consisting of two heavy
quarks (b,c), the $B_c$ meson is usually treated as a heavy
quarkonium system.  In the non-relativistic limit, the $B_c$ wave
function can be written as \cite{prd81014022}
\begin{eqnarray}
\Phi_{B_c}(x)=\frac{if_B}{4N_c}[(\rlap{/}{P}+M_{B_c})\gamma_5\delta(x-r_c)],
\end{eqnarray}
with $r_c=m_c/M_{B_c}$. Here, we only consider one of the dominant
Lorentz structure, and neglect another contribution in our
calculation \cite{epjc28515}.

 In the heavy quark limit, the two-particle light-cone distribution
amplitudes of $D_{(s)}/D_{(s)}^*$ meson are defined
as \cite{prd67054028}
\begin{eqnarray}\label{eq:dwave}
\langle D_{(s)}(P_2)|q_{\alpha}(z)\bar{c}_{\beta}(0)|0\rangle &=&
\frac{i}{\sqrt{2N_c}}\int^1_0dx e^{ixP_2\cdot
z}[\gamma_5(\rlap{/}{P}_2+m_{D_{(s)}})\phi_{D_{(s)}}(x,b)]_{\alpha\beta},
\nonumber\\
\langle D_{(s)}^*(P_2)|q_{\alpha}(z)\bar{c}_{\beta}(0)|0\rangle
&=&-\frac{1}{\sqrt{2N_c}}\int^1_0dx e^{ixP_2\cdot
z}[\rlap{/}{\epsilon} (\rlap{/}{P}_2+m_{D_{(s)}^*})\phi
_{D_{(s)}^*}(x,b)
]_{\alpha\beta} .
\end{eqnarray}

We use the following relations derived from HQET to determine
$f_{D^*_{(s)}}$ \cite{hqet}
\begin{eqnarray}
f_{D^*_{(s)}}=\sqrt{\frac{m_{D_{(s)}}}{m_{D_{(s)}^*}}}f_{D_{(s)}}.
\end{eqnarray}
For the $D_{(s)}^{(*)}$ meson wave function, we adopt the same model
as of the $B$ meson \cite{epjc45711}
\begin{eqnarray}
\phi_{D_{(s)}}(x,b)=N_{D_{(s)}}[x(1-x)]^2\exp\left(-\frac{x^2m_{D_{(s)}}^2}{2\omega_{D_{(s)}}^2}-\frac{1}{2}\omega_{D_{(s)}}^2b^2\right)
\end{eqnarray}
with       shape parameters  $\omega_D=0.6$ for $D/D^*$ meson and
$\omega_{D_s}=0.8$ for $D_s/D_s^*$ meson. Here, a larger
$\omega_{D_s}$ parameter than $\omega_{D}$  characterize the fact
that the  s quark in $D_s$ meson carries a larger momentum fraction
than the light quark (u,d) in the $D$ meson.

\begin{figure}[htbh]
\begin{center}
\vspace{-1cm} \centerline{\epsfxsize=12 cm \epsffile{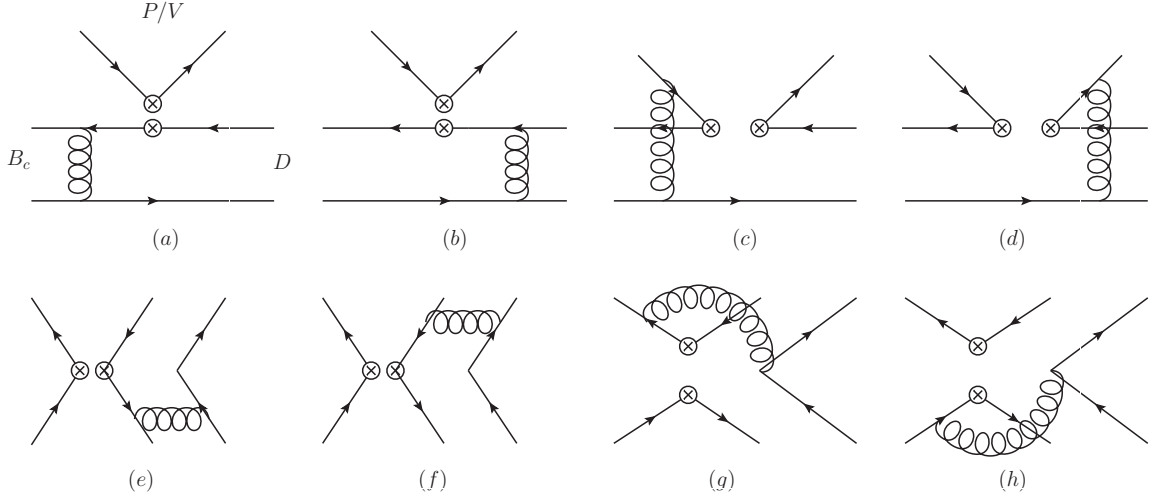}}
\vspace{-8cm} \caption{The leading order Feynman diagrams for
the decays  $B_c\rightarrow D^{(*)}_{(s)}P,D^{(*)}_{(s)}V$. }
 \label{fig:lodiagram}
 \end{center}
\end{figure}

At leading order, there are eight types of diagrams that may
contribute to the $B_c\rightarrow D_{(s)}^{(*)}P$, $D_{(s)}^{(*)}V$ decays as
illustrated in Fig.\ref{fig:lodiagram}. The first line are the
emission  type diagrams, with the first two contributing to the
usual form factor; the last two so-called non-factorizable diagrams.
The second line are the annihilation type diagrams, with the first
two factorizable; the last two non-factorizable.

\subsection{ Amplitudes for $B_c\rightarrow D_{(s)}P$ decays}\label{sec:bcdp}

We mark LL, LR, SP to denote the contributions from $(V-A)(V-A)$, $(V-A)(V+A)$ and $(S-P)(S+P)$ operates, respectively.
The amplitudes from factorizable diagrams (a) and (b) in  Fig.\ref{fig:lodiagram} are as following:
\begin{eqnarray}\label{eq:ll}
\mathcal {F}^{LL}_e&=&2\sqrt{\frac{2}{3}}C_ff_Bf_{P}\pi M_B^4
\int_0^1dx_2\int_0^{\infty}b_1b_2db_1db_2\phi_{D}(x_2,b_2)\times\nonumber\\&&\{
[(1-2r_{D})x_2+(r_D-2)r_b]\alpha_s(t_a)h_e(\alpha_e,\beta_a,b_1,b_2)S_t(x_2)\exp[-S_{ab}(t_a)]
\nonumber\\&&-(r_D-2)r_D(x_1-1)\alpha_s(t_b)h_e(\alpha_e,\beta_b,b_2,b_1)S_t(x_1)\exp[-S_{ab}(t_b)],
\end{eqnarray}
where $r_D=m_D/M_B$, $r_b=m_b/M_B$; $C_F=4/3$ is a color factor;
$f_P$ is the decay constant of pseudoscalar meson (P). The
factorization scales $t_{a,b}$ are chosen as the maximal virtuality
of internal particles in the hard amplitude, in order to suppress
the higher order corrections \cite{prd074004}. The function $h_e$
 are displayed in the Appendix \ref{sec:b}. The factor
$S_t(x)$ is the jet function from the threshold resummation,
 whose definitions can be found in \cite{epjc45711}.
The terms proportional to $r_D^2$ have been neglected for small values. We
can calculate the form factor from eq.(\ref{eq:ll}) if we take
away the Wilson coefficients and $f_P$.
For the $(V-A)(V+A)$ operates, we have $\mathcal {F}^{LR}_e=-\mathcal {F}^{LL}_e$ since only axial-vector current
contributes to the pseudoscalar meson production.
For the $(S-P)(S+P)$ operates the formula is different:
\begin{eqnarray}
\mathcal {F}^{SP}_e&=&-4\sqrt{\frac{2}{3}}C_ff_Bf_{P}\pi M_B^4
\int_0^1dx_2\int_0^{\infty}b_1b_2db_1db_2\phi_{D}(x_2,b_2)\times\nonumber\\&&\{
[r_D(4r_b-x_2-1)-r_b+2]\alpha_s(t_a)h_e(\alpha_e,\beta_a,b_1,b_2)S_t(x_2)\exp[-S_{ab}(t_a)]
\nonumber\\&&+[r_D(2-4x_1)+x_1]\alpha_s(t_b)h_e(\alpha_e,\beta_b,b_2,b_1)S_t(x_1)\exp[-S_{ab}(t_b)].
\end{eqnarray}

For the nonfactorizable emission diagram (c) and (d), the decay amplitudes of
three types operates are:
\begin{eqnarray}
\mathcal {M}_e^{LL}&=&\frac{8}{3}C_Ff_B\pi M_B^4
\int_0^1dx_2dx_3\int_0^{\infty}b_2b_3db_2db_3\phi_{D}(x_2,b_2)\phi_P^A(x_3)\times\nonumber\\&&
\{[r_D(1-x_1-x_2)+x_1+x_3-1]\alpha_s(t_c)h_e(\beta_c,\alpha_e,b_3,b_2)\exp[-S_{cd}(t_c)]-
\nonumber\\&&[r_D(1-x_1-x_2)+2x_1+x_2-x_3-1]\alpha_s(t_d)h_e(\beta_d,\alpha_e,b_3,b_2)\exp[-S_{cd}(t_d)]
\},\nonumber\\&&
\end{eqnarray}
\begin{eqnarray}
\mathcal {M}_e^{LR}&=&\frac{8}{3}C_Ff_B\pi M_B^4 r_P(1+r_D)
\int_0^1dx_2dx_3\int_0^{\infty}b_2b_3db_2db_3\phi_{D}(x_2,b_2)\times\nonumber\\&&
\{[(x_1+x_3-1+r_D(2x_1+x_2+x_3-2))\phi_P^P(x_3)+\nonumber\\&&
(x_1+x_3-1+r_D(x_3-x_2))\phi_P^T(x_3)]
\alpha_s(t_c)h_e(\beta_c,\alpha_e,b_3,b_2)\exp[-S_{cd}(t_c)]
\nonumber\\&&-[(x_1-x_3+r_D(2x_1+x_2-x_3-1))\phi_P^P(x_3)+(x_3-x_1+\nonumber\\&&
r_D(x_3+x_2-1))\phi_P^T(x_3)]
\alpha_s(t_d)h_e(\beta_d,\alpha_e,b_3,b_2)\exp[-S_{cd}(t_d)]
\},
\end{eqnarray}
\begin{eqnarray}
\mathcal {M}_e^{SP}&=&\frac{8}{3}C_Ff_B\pi M_B^4
\int_0^1dx_2dx_3\int_0^{\infty}b_2b_3db_2db_3\phi_{D}(x_2,b_2)\phi_P^A(x_3)\times\nonumber\\&&
\{[r_D(x_1+x_2-1)-2x_1-x_2-x_3+2]\nonumber\\&&\alpha_s(t_c)h_e(\beta_c,\alpha_e,b_3,b_2)\exp[-S_{cd}(t_c)]-
\nonumber\\&&[x_3-x_1-r_D(1-x_1-x_2)]\alpha_s(t_d)h_e(\beta_d,\alpha_e,b_3,b_2)\exp[-S_{cd}(t_d)]
\},
\end{eqnarray}
where $r_P=m_0^P/M_{B}$, with $m_0^P$ as the chiral mass of
the pseudoscalar meson P.

  For the factorizable emission diagram (e) and (f), we keep the mass of the c-quark in $D$ meson,  while the
mass of the light quark is neglected. The amplitudes are given
as follows:
\begin{eqnarray}
\mathcal {F}_a^{LL}&=&\mathcal {F}_a^{LR}=-8C_Ff_B\pi M_B^4
\int_0^1dx_2dx_3\int_0^{\infty}b_2b_3db_2db_3\phi_{D}(x_2,b_2)\times\nonumber\\&&
\{[\phi_P^A(x_3)(x_3-2r_Dr_c)+r_P[\phi_P^P(x_3)(2r_D(x_3+1)-r_c)+\phi_P^T(x_3)\nonumber\\&&
(r_c+2r_D(x_3-1))]]\alpha_s(t_e)h_e(\alpha_a,\beta_e,b_2,b_3)\exp[-S_{ef}(t_e)]S_t(x_3)-
\nonumber\\&&[x_2\phi_P^A(x_3)+2r_Pr_D(x_2+1)\phi_P^P(x_3)]\nonumber\\&&
\alpha_s(t_f)h_e(\alpha_a,\beta_f,b_3,b_2)\exp[-S_{ef}(t_f)]S_t(x_2)
\},
\end{eqnarray}
\begin{eqnarray}
\mathcal {F}_a^{SP}&=&16C_Ff_B\pi M_B^4
\int_0^1dx_2dx_3\int_0^{\infty}b_2b_3db_2db_3\phi_{D}(x_2,b_2)\times\nonumber\\&&
\{[-\phi_P^A(x_3)(2r_D-r_c)+r_P[\phi_P^P(x_3)(4r_cr_D-x_3)+\phi_P^T(x_3)x_3]]\nonumber\\&&
\alpha_s(t_e)h_e(\alpha_a,\beta_e,b_2,b_3)\exp[-S_{ef}(t_e)]S_t(x_3)-[x_2r_D\phi_P^A(x_3)+
\nonumber\\&&2r_P\phi_P^P(x_3)]\alpha_s(t_f)h_e(\alpha_a,\beta_f,b_3,b_2)\exp[-S_{ef}(t_f)]S_t(x_2)
\};
\end{eqnarray}
and that of the nonfactorizable annihilation diagram (g) and (h) are
\begin{eqnarray}
\mathcal {M}_a^{LL}&=&-\frac{8}{3}C_Ff_B\pi M_B^4
\int_0^1dx_2dx_3\int_0^{\infty}b_1b_2db_1db_2\phi_{D}(x_2,b_2)\times\nonumber\\&&
\{[\phi_P^A(x_3)(r_c-x_1+x_2)+r_Pr_D[\phi_P^T(x_3)(x_2-x_3)
+\nonumber\\&&\phi_P^P(x_3)(4r_c-2x_1+x_2+x_3)]]\alpha_s(t_g)h_e(\beta_g,\alpha_a,b_1,b_2)\exp[-S_{gh}(t_g)]\nonumber\\&&+
[-\phi_P^A(x_3)(r_b+x_1+x_3-1)+r_Pr_D[(x_2-x_3)\phi_P^T(x_3)-\phi_P^P(x_3)\nonumber\\&&
(4r_b+2x_1+x_2+x_3-2)]]
\alpha_s(t_h)h_e(\beta_h,\alpha_a,b_1,b_2)\exp[-S_{gh}(t_h)]
\},
\end{eqnarray}
\begin{eqnarray}
\mathcal {M}_a^{LR}&=&\frac{8}{3}C_Ff_B\pi M_B^4
\int_0^1dx_2dx_3\int_0^{\infty}b_1b_2db_1db_2\phi_{D}(x_2,b_2)\times\nonumber\\&&
\{[-\phi_P^A(x_3)r_D(r_c+x_1-x_2)+r_P[-\phi_P^T(x_3)(-r_c-x_1+x_3)
\nonumber\\&&+\phi_P^P(x_3)(r_c+x_1-x_3)]]\alpha_s(t_g)h_e(\beta_g,\alpha_a,b_1,b_2)\exp[-S_{gh}(t_g)]+
\nonumber\\&&[-\phi_P^A(x_3)r_D(-r_b+x_1+x_2-1)+r_P[(-r_b+x_1+x_3-1)\nonumber\\&&(\phi_P^P(x_3)+
\phi_P^T(x_3))]]
\alpha_s(t_h)h_e(\beta_h,\alpha_a,b_1,b_2)\exp[-S_{gh}(t_h)]
\},
\end{eqnarray}
\begin{eqnarray}\label{eq:llend}
\mathcal {M}_a^{SP}&=&-\frac{8}{3}C_Ff_B\pi M_B^4
\int_0^1dx_2dx_3\int_0^{\infty}b_1b_2db_1db_2\phi_{D}(x_2,b_2)\times\nonumber\\&&
\{[-\phi_P^A(x_3)(x_1-x_3-r_c)+r_Pr_D[-\phi_P^T(x_3)(x_2-x_3)
\nonumber\\&&+\phi_P^P(x_3)(4r_c-2x_1+x_2+x_3)]]\alpha_s(t_g)h_e(\beta_g,\alpha_a,b_1,b_2)\exp[-S_{gh}(t_g)]+
\nonumber\\&&[-\phi_P^A(x_3)(r_b+x_1+x_2-1)+r_Pr_D[(-4r_b-2x_1-x_2-x_3+2)\phi_P^P(x_3)\nonumber\\&&-(x_2-x_3)
\phi_P^T(x_3))]]
\alpha_s(t_h)h_e(\beta_h,\alpha_a,b_1,b_2)\exp[-S_{gh}(t_h)]
\},
\end{eqnarray}
With the functions obtained in the above,
the total decay amplitudes of 10 decay channels for the $B_c\rightarrow D_{(s)}P$  can be given by
\begin{eqnarray}\label{eq:dpi}
\mathcal {A}(B_c\rightarrow D^0\pi^+)&=&\xi_u[a_1\mathcal {F}_e^{LL}+C_1\mathcal {M}_e^{LL}]
+\xi_c[a_1\mathcal {F}_a^{LL}+C_1\mathcal {M}_a^{LL}]\nonumber\\&&
-\xi_t[(C_3+C_9)(\mathcal {M}_e^{LL}+\mathcal {M}_a^{LL})+(C_5+C_7)(\mathcal {M}_e^{LR}+\mathcal {M}_a^{LR})\nonumber\\&&
+(C_4+\frac{1}{3}C_3+C_{10}+\frac{1}{3}C_9)(\mathcal {F}_a^{LL}+\mathcal {F}_e^{LL})\nonumber\\&&
+(C_6+\frac{1}{3}C_5+C_{8}+\frac{1}{3}C_7)(\mathcal {F}_a^{SP}+\mathcal {F}_e^{SP})],
\end{eqnarray}
\begin{eqnarray}
\sqrt{2}\mathcal {A}(B_c\rightarrow D^+\pi^0)&=&\xi_u[a_2\mathcal {F}_e^{LL}+C_2\mathcal {M}_e^{LL}]
-\xi_c[a_1\mathcal {F}_a^{LL}\nonumber\\&&+C_1\mathcal {M}_a^{LL}]
-\xi_t[(\frac{3}{2}C_{10}-C_3+\frac{1}{2}C_9)\mathcal {M}_e^{LL}-(C_3+C_9)\mathcal {M}_a^{LL}
\nonumber\\&&+(-C_5+\frac{1}{2}C_7)\mathcal {M}_e^{LR}+
(-C_4-\frac{1}{3}C_3-C_{10}-\frac{1}{3}C_9)\mathcal {F}_a^{LL}+\nonumber\\&&
(C_{10}+\frac{5}{3}C_9-\frac{1}{3}C_3-C_4-\frac{3}{2}C_7-\frac{1}{2}C_8)\mathcal {F}_e^{LL}
\nonumber\\&&+(-C_6-\frac{1}{3}C_5+\frac{1}{2}C_{8}+\frac{1}{6}C_7)\mathcal {F}_e^{SP}-(C_5+C_7)\mathcal {M}_a^{LR}
\nonumber\\&&+(-C_6-\frac{1}{3}C_5-C_{8}-\frac{1}{3}C_7)\mathcal {F}_a^{SP}],
\end{eqnarray}
\begin{eqnarray}\label{eq:detaq}
\sqrt{2}\mathcal {A}(B_c\rightarrow D^+\eta_q)&=&\xi_u[a_2\mathcal {F}_{e}^{LL}+C_2\mathcal {M}_{e}^{LL}]
+\xi_c[a_1\mathcal {F}_{a}^{LL}+C_1\mathcal {M}_{a}^{LL}]\nonumber\\&&
-\xi_t[(2C_4+C_3+\frac{1}{2}C_{10}-\frac{1}{2}C_9)\mathcal {M}_{e}^{LL}+(C_3+C_9)\mathcal {M}_{a}^{LL}\nonumber\\&&
+(C_5-\frac{1}{2}C_7)\mathcal {M}_{e}^{LR}+(C_5+C_7)\mathcal {M}_{a}^{LR}+
(C_4+\frac{1}{3}C_3+C_{10}+\nonumber\\&&\frac{1}{3}C_9)\mathcal {F}_{a}^{LL}+
(\frac{7}{3}C_3+\frac{5}{3}C_4+\frac{1}{3}(C_9-C_{10}))\mathcal {F}_{e}^{LL}+
(2C_5+\frac{2}{3}C_6\nonumber\\&&+\frac{1}{2}C_7+\frac{1}{6}C_8)\mathcal {F}_{e}^{LR}+
(C_6+\frac{1}{3}C_5-\frac{1}{2}C_{8}-\frac{1}{6}C_7)\mathcal {F}_{e}^{SP}\nonumber\\&&+
(C_6+\frac{1}{3}C_5+C_{8}+\frac{1}{3}C_7)\mathcal {F}_{a}^{SP}],
\end{eqnarray}
\begin{eqnarray}\label{eq:detas}
\mathcal {A}(B_c\rightarrow D^+\eta_s)&=&-\xi_t[(C_4-\frac{1}{2}C_{10})\mathcal {M}_{e}^{LL}
+(C_6-\frac{1}{2}C_8)\mathcal {M}_{e}^{SP}+
(C_3+\frac{1}{3}C_4\nonumber\\&&-\frac{1}{2}C_9-\frac{1}{6}C_{10})\mathcal {F}_{e}^{LL}+(C_5+\frac{1}{3}C_6-\frac{1}{2}C_7-\frac{1}{6}C_8)\mathcal {F}_{e}^{LR}],
\end{eqnarray}
\begin{eqnarray}\label{eq:dsetaq}
\sqrt{2}\mathcal {A}(B_c\rightarrow D_s^+\eta_q)&=&\xi'_u[a_2\mathcal {F}_{e}^{LL}+C_2\mathcal {M}_{e}^{LL}]
-\xi'_t[(2C_4+\frac{1}{2}C_{10})\mathcal {M}_{e}^{LL}+\nonumber\\&&(2C_6+\frac{1}{2}C_8)\mathcal {M}_{e}^{SP}+
(2C_3+\frac{2}{3}C_4+\frac{1}{2}C_9+\frac{1}{6}C_{10})\mathcal {F}_{e}^{LL}\nonumber\\&&+(2C_5+\frac{2}{3}C_6+\frac{1}{2}C_7+\frac{1}{6}C_8)\mathcal {F}_{e}^{LR}],
\end{eqnarray}
\begin{eqnarray}\label{eq:dsetas}
\mathcal {A}(B_c\rightarrow D_s^+\eta_s)&=&
\xi'_c[a_1\mathcal {F}_{a}^{LL}+C_1\mathcal {M}_{a}^{LL}]
-\xi'_t[
(C_3+C_4-\frac{1}{2}(C_{10}+C_9))\mathcal {M}_{e}^{LL}\nonumber\\&&+(C_3+C_9)\mathcal {M}_{a}^{LL}
+(C_5-\frac{1}{2}C_7)\mathcal {M}_{e}^{LR}+(C_5+C_7)\mathcal {M}_{a}^{LR}\nonumber\\&&+
(C_4+\frac{1}{3}C_3+C_{10}+\frac{1}{3}C_9)\mathcal {F}_{a}^{LL}+(C_6-\frac{1}{2}C_8)\mathcal {M}_{e}^{SP}
+\nonumber\\&&
\frac{2}{3}(2(C_3+C_4)-(C_9+C_{10}))\mathcal {F}_{e}^{LL}+(C_5+\frac{1}{3}C_6-\frac{1}{2}C_7-\frac{1}{6}C_8)\mathcal {F}_{e}^{LR}\nonumber\\&&+
(C_6+\frac{1}{3}C_5-\frac{1}{2}C_{8}-\frac{1}{6}C_7)\mathcal {F}_{e}^{SP}+
(C_6+\frac{1}{3}C_5+C_{8}+\frac{1}{3}C_7)\mathcal {F}_{a}^{SP}],\nonumber\\&&
\end{eqnarray}
\begin{eqnarray}\label{eq:dk}
\mathcal {A}(B_c\rightarrow D_s^+\bar{K}^0)&=&
\xi_c[a_1\mathcal {F}_a^{LL}+C_1\mathcal {M}_a^{LL}]
-\xi_t[(C_3-\frac{1}{2}C_9)\mathcal {M}_e^{LL}\nonumber\\&&+(C_3+C_9)\mathcal {M}_a^{LL}
+(C_5-\frac{1}{2}C_7)\mathcal {M}_e^{LR}+(C_5+C_7)\mathcal {M}_a^{LR}\nonumber\\&&+
(C_4+\frac{1}{3}C_3+C_{10}+\frac{1}{3}C_9)\mathcal {F}_a^{LL}+\nonumber\\&&
(C_{4}+\frac{1}{3}C_3-\frac{1}{2}C_{10}-\frac{1}{6}C_9)\mathcal {F}_e^{LL}
+(C_6+\frac{1}{3}C_5-\frac{1}{2}C_{8}\nonumber\\&&-\frac{1}{6}C_7)\mathcal {F}_{e}^{SP}
+(C_6+\frac{1}{3}C_5+C_{8}+\frac{1}{3}C_7)\mathcal {F}_a^{SP}],
\end{eqnarray}
\begin{eqnarray}\label{eq:dpi0}
\mathcal {A}(B_c\rightarrow D_s^+ \pi^0)&=&\xi'_u[a_2\mathcal {F}_e^{LL}+C_2\mathcal {M}_e^{LL}]
-\xi'_t[(\frac{1}{2}(3C_9+C_{10})\mathcal {F}_e^{LL}\nonumber\\&&+\frac{1}{2}(3C_7+C_8))\mathcal {F}_e^{LR}
+\frac{3}{2}C_{10}\mathcal {M}_e^{LL}+\frac{3}{2}C_{8}\mathcal {M}_e^{SP}],
\end{eqnarray}
with the CKM matrix element
$\xi_{i}=V_{id}V^*_{ib}$ and $\xi'_{i}=V_{is}V^*_{ib}(i=u,c,t)$. The
combinations  Wilson coefficients $a_1=C_2+C_1/3$ and  $a_2=C_1+C_2/3$.
The total decay amplitude of $\mathcal {A}(B_c\rightarrow D^0 K^+)$ and $\mathcal {A}(B_c\rightarrow D^+ K^0)$  can be  obtained from (\ref{eq:dpi})
and (\ref{eq:dk}), respectively,
with the following  replacement:
\begin{eqnarray}\label{eq:dkk0}
\mathcal {A}(B_c\rightarrow D^0 K^+)&=&\mathcal {A}(B_c\rightarrow D^0 \pi^+)|_{\pi\rightarrow K, \xi_{i}\rightarrow \xi'_{i} },\nonumber\\
\mathcal {A}(B_c\rightarrow D^+ K^0)&=&\mathcal {A}(B_c\rightarrow D_s^+\bar{K}^0)|_{D_s\rightarrow D, \xi_{i}\rightarrow \xi'_{i} }.
\end{eqnarray}
It should be noticed that, in (\ref{eq:detaq}), (\ref{eq:detas}), (\ref{eq:dsetaq}) and (\ref{eq:dsetas}),
 the decay amplitudes are for the mixing basis of $(\eta_q, \eta_s)$.
 For the physical state $(\eta, \eta')$,
 the decay amplitudes are
\begin{eqnarray}\label{eq:detaetap}
\mathcal {A}(B_c\rightarrow D^+ \eta)&=&\mathcal {A}(B_c\rightarrow D^+ \eta_q)\cos \phi -\mathcal {A}(B_c\rightarrow D^+ \eta_s)\sin \phi,\nonumber\\
\mathcal {A}(B_c\rightarrow D^+ \eta')&=&\mathcal {A}(B_c\rightarrow D^+ \eta_q) \sin\phi +\mathcal {A}(B_c\rightarrow D^+ \eta_s)\cos \phi,\nonumber\\
\mathcal {A}(B_c\rightarrow D_s^+ \eta)&=&\mathcal {A}(B_c\rightarrow D_s^+ \eta_q)\cos \phi -\mathcal {A}(B_c\rightarrow D_s^+ \eta_s)\sin \phi,\nonumber\\
\mathcal {A}(B_c\rightarrow D_s^+ \eta')&=&\mathcal {A}(B_c\rightarrow D_s^+ \eta_q) \sin\phi +\mathcal {A}(B_c\rightarrow D_s^+ \eta_s)\cos \phi,
\end{eqnarray}
where $\phi=39.3^{\circ}$ is the mixing angle
between the two states.
\begin{eqnarray}
\left(
\begin{array}{c}
\eta\\
\eta'\\
\end{array}
\right)=\left(
\begin{array}{cc}
\cos \phi & -\sin \phi\\
\sin \phi & \cos \phi\\
\end{array}
\right)\left(
\begin{array}{c}
\eta_q\\
\eta_s\\
\end{array}
\right).
\end{eqnarray}

\subsection{ Amplitudes for $B_c\rightarrow D_{(s)}V$ decays}\label{sec:bcdv}

In $B_c\rightarrow D_{(s)}V$ decays, the vector meson is longitudinally polarized. In the leading power contribution,
the formula of each Feynman diagram for $B_c\rightarrow D_{(s)}V$ is similar to that of the  $B_c\rightarrow D_{(s)}P$  modes, but with the replacements
\begin{eqnarray}
f_P\rightarrow f_V,\quad r_P\rightarrow r_V,\quad \phi_P^A\rightarrow \phi_V,\quad  \phi_P^P\rightarrow -\phi_V^s, \quad  \phi_P^T\rightarrow \phi_V^t.
\end{eqnarray}
 The total decay amplitude
for $B_c\rightarrow D_{(s)}V$ can be obtained through the substitutions in (\ref{eq:dpi})- (\ref{eq:dkk0}):
\begin{eqnarray}
\pi\rightarrow \rho,\quad K\rightarrow K^*,\quad \eta_q\rightarrow \omega ,\quad \eta_s\rightarrow \phi.
\end{eqnarray}

\subsection{ Amplitudes for $B_c\rightarrow D^*_{(s)}V$ decays}\label{sec:bcdsv}

The decay amplitude of $B_c\rightarrow D^*_{(s)}V$ can be decomposed into
\begin{eqnarray}
\mathcal {A}(\epsilon_{D^*},\epsilon_V)=\mathcal {A}^L+\mathcal {A}^N \epsilon_{D^*T} \cdot \epsilon_{VT}+
i \mathcal {A}^T \epsilon_{\alpha\beta\rho\sigma}n^{\alpha}v^{\beta}\epsilon_{D^*T}^{\rho}\epsilon_{VT}^{\sigma}
\end{eqnarray}
where $\epsilon_{D^*T}(\epsilon_{VT})$ is the transverse polarization vector for $D^*(V)$ meson.
$\mathcal {A}^L$ corresponds to the contributions of  longitudinal polarization; $\mathcal {A}^N$ and $\mathcal {A}^T$
corresponds to the contributions of  normal  and transverse  polarization, respectively.
 The factorization formulae for the longitudinal, normal and transverse polarizations  are
all listed in Appendix \ref{sec:a}.
There are also 10 channels for  $B_c\rightarrow D^*_{(s)}V$ decay modes. We can obtain the total decay amplitudes from those in
$B_c\rightarrow D_{(s)}V$  with replacing $D_{(s)}$ by $D_{(s)}^*$.

\subsection{ Amplitudes for $B_c\rightarrow D^*_{(s)}P$ decays}\label{sec:bcdsp}

For $B_c\rightarrow D^*_{(s)}P$, only the  longitudinal polarization of  $D^*_{(s)}$
will contribute. We can obtain their amplitudes  from the longitudinal polarization amplitudes for the
$B_c\rightarrow D^*_{(s)}V$ decays with the following replacement in the distribution amplitudes:
\begin{eqnarray}
f_V\rightarrow f_P,\quad r_V\rightarrow r_P,\quad \phi_V\rightarrow \phi^A_P, \quad \phi^s_V\rightarrow \phi^P_P,\quad \phi^t_V\rightarrow \phi^T_P.
\end{eqnarray}
In fact,   the $B_c\rightarrow D^*_{(s)}P$ decays amplitude  are the same as the $B_c\rightarrow D_{(s)}P$ ones only at
leading power under the hierachy $M_{B_c}\gg m_{D^{(*)}}\gg \Lambda_{QCD}$.
An explicit derivation shows
that the difference  between the two kinds of channels occurs at $\mathcal {O}(r_{D^{(*)}})$  and at
the twist-3 level in eq.(\ref{eq:ll})-eq.(\ref{eq:llend}).

\section{NUMERICAL results AND DISCUSSIONS }\label{sec:numerical}


The numerical results of   our calculations depend on a set of input
parameters. We list the decay constants of various mesons and
parameters of hadronic wave functions in Table \ref{tab:constant}.
For $\eta-\eta'$ system, the decay constants $f_q$ and $f_s$ in the
quark-flavor basis have been extracted from various related
experiments \cite{prd58114006,prd48339}
\begin{eqnarray}
f_q=(1.07\pm 0.02)f_{\pi},\quad f_s=(1.34\pm 0.06)f_{\pi}.
\end{eqnarray}
For the CKM matrix elements, the quark masses etc., we adopt the
results from \cite{npp37}
\begin{eqnarray}\label{eq:ckm}
|V_{ub}|&=&(3.89\pm 0.44)\times 10^{-3},\quad |V_{ud}|=0.97425,\quad |V_{cb}|=0.0406,\quad |V_{cd}|=0.23
\nonumber\\  |V_{us}|&=&0.2252,\quad |V_{cs}|=1.023, \quad \gamma=(73^{+22}_{-25})^{\circ},
\nonumber\\
m_c&=&1.27 \text{GeV},\quad m_b=4.2\text{GeV},
\quad m_0^{\pi}=1.4\text{GeV},\nonumber\\
 m_0^{K}&=&1.6\text{GeV},\quad m_0^{\eta_q}=1.07\text{GeV},\quad m_0^{\eta_s}=1.92\text{GeV},
 \quad \Lambda^5_{QCD}=0.112\text{GeV}.
\end{eqnarray}

\begin{table}
\caption{The decay constants and the hadronic  meson wave function
  parameters taken from the light-cone sum
rules \cite{prd71014015}.} \label{tab:constant}
\begin{tabular*}{12.1cm}{@{\extracolsep{\fill}}|l|c|c|c|c|c|c|c|c|c|c|c|c|c|c|}
  \toprule[2pt]
  \multicolumn{13}{|c|}{ The decay constants (MeV)}\\ \hline
$f_{B_c}$& $f_{D}$& $f_{D_s}$& $f_{\pi}$ & $f_{K}$
&$f_{\rho}$&$f^T_{\rho}$&$f_{\omega}$&$f^T_{\omega}$
&$f_{\phi}$&$f^T_{\phi}$&$f_{K^*}$&$f^T_{K^*}$ \\ \hline
489&$206.7\pm 8.9$&$257.5\pm 6.1$&131&160 &209 &165&195&145&231&200&217&185 \\
\bottomrule[2pt]
\end{tabular*}
\begin{tabular*}{12.1cm}{@{\extracolsep{\fill}}|l|l|l|l|l|}
  \toprule[2pt]
  \multicolumn{5}{|c|}{ Values of Gegenbauer moments}\\ \hline
 &$\pi$ & $K$ & $\eta_q$ & $\eta_s$ \\\hline
$a^P_1$&--&$0.17$&--&--  \\\hline
$a^P_2$&$0.25$&$0.115$&$0.115$&$0.115$  \\\hline
$a^P_4$&-0.015&-0.015&-0.015&-0.015 \\\hline
  &$\rho$& $\omega$ &$\phi$ &$K^*$ \\\hline
$a^{\parallel}_1$&--&--&-- &$0.03$  \\\hline
$a^{\parallel}_2$&$0.15$&$0.15$&$0.18$ &$0.11$  \\\hline
$a^{\perp}_1$&--&--&-- &$0.04$  \\\hline
$a^{\perp}_2$&$0.14$&$0.14$&$0.14$ &$0.10$  \\\bottomrule[2pt]
\end{tabular*}
\end{table}


For the considered $B_c\rightarrow D_{(s)}P$,  $B_c\rightarrow D^*_{(s)}P$ and $B_c\rightarrow D_{(s)}V$ decays, the branching ratios $\mathcal {BR}$
and the direct CP asymmetry $A^{dir}_{CP}$ for a
given mode can be written as
\begin{eqnarray}
\mathcal {BR}=\frac{G_F\tau_{B_c}}{32\pi M_B}(1-r_D^2)|\mathcal {A}|^2,\quad
A^{dir}_{CP}=\frac{|\mathcal {\bar{A}}|^2-|\mathcal {A}|^2}{|\mathcal {\bar{A}}|^2+|\mathcal {A}|^2}
\end{eqnarray}
where the  decay amplitudes $\mathcal {A}$ have been given
explicitly in Sec. \ref{sec:f-work} for each channel. $\mathcal
{\bar{A}}$ is the corresponding charge conjugate decay amplitude,
which can be obtained by conjugating the CKM matrix elements in
$\mathcal {A}$.

 Our numerical results of CP averaged branching ratios  and direct CP asymmetries for $ B_c\rightarrow
 D_{(s)}P$ and $ B_c\rightarrow
 D_{(s)}V$   decays
 are listed
 in Tables \ref{tab:bcdp} and \ref{tab:bcdv}, respectively.
The dominant topologies contributing to these decays are also indicated through the symbols
T (color-allowed tree), C (color-suppressed tree), P (penguin) and A (annihilation).
 The first theoretical error in
 all our tables
  is referred to the {\bf $D_{(s)}$ } meson wave function: (1) The
shape parameter $\omega_{D}= 0.60 \pm 0.05$ for $D/D^*$ meson and
$\omega_{D_{s}}= 0.80 \pm 0.05$ for $D_s/D_s^*$ meson;  (2) The decay constant
$f_{D}= (206.7 \pm 8.9) \text{MeV}$ for $D$ meson and $f_{D_s}= (257.5 \pm 6.1) \text{MeV}$ for $D_s$ meson.
The second error is from the combined
uncertainty in the CKM matrix elements $V_{ub}$  and
the angle of unitarity triangle $\gamma$,  which are given in eq.(\ref{eq:ckm}). The third error arises from the
hard scale t varying from $0.75t$ to $1.25t$, which characterizing
the size of next-to-leading order QCD contributions.
Most of the branching ratios are  sensitive to the hadronic
parameters and the CKM matrix elements. The CP asymmetry
parameter is only sensitive to the next-to-leading order   contributions, since the uncertainty of hadronic parameters are mostly canceled by the ratios.

\begin{table}
\caption{ CP averaged branching ratios  and direct CP asymmetries
for $ B_c\rightarrow D_{(s)}P$  decays, together with results from RCQM and LFQM.} \label{tab:bcdp}
\begin{tabular}[t]{l|c|c|c|c|c|c}
\toprule[2pt]
\multicolumn{2}{c|}{ } &\multicolumn{3}{c}{ $\mathcal {BR}(10^{-7})$} &\multicolumn{2}{|c}{ $A^{dir}_{CP}(\%)$}  \\ \hline
 channels &Class & This work & RCQM\footnotemark[1] & LFQM & This work & RCQM \\ \hline
$B_c\rightarrow D^0\pi^+$ &T& $26.7^{+3.1+6.0+0.8}_{-3.5-5.6-0.6}$ &22.9&4.3 &$-41.2^{+4.5+11.1+0.8}_{-4.6-7.8-1.2}$ &6.5\\
$B_c\rightarrow D^+\pi^0$ &C,A & $0.82^{+0.24+0.55+0.06}_{-0.16-0.41-0.01}$&2.1&0.067&$2.3^{+6.3+1.4+15.0}_{-3.0-0.8-18.8}$&-1.9\\
$B_c\rightarrow D^0K^+$ &A,P & $47.8^{+17.2+2.2+5.4}_{-9.1-1.7-3.6}$&44.5&0.35&$-34.8^{+4.9+7.4+1.8}_{-2.6-3.7-1.3}$&-4.6\\
$B_c\rightarrow D^+K^0$ &A,P& $46.9^{+15.6+0.3+7.4}_{-12.3-0.3-4.6}$&49.3&--&$2.3^{+0.4+0.9+0.0}_{-0.2-0.5-0.0}$&-0.8\\
$B_c\rightarrow D^+\eta$ &C,A&$0.92^{+0.15+0.21+0.03}_{-0.15-0.25-0.00}$&--&0.087&$40.8^{+0.0+18.4+15.6}_{-2.9-14.0-13.5}$&--\\
$B_c\rightarrow D^+\eta'$ &C,A&$0.91^{+0.12+0.16+0.06}_{-0.10-0.20-0.03}$&--&0.048&$-14.0^{+0.6+4.6+15.9}_{-1.5-5.2-11.9}$&--\\
$B_c\rightarrow D_s^+\pi^0$ &C,P&$0.41^{+0.04+0.01+0.02}_{-0.04-0.02-0.02}$&--&0.0067&$46.7^{+1.4+6.3+2.5}_{-1.4-11.8-2.8}$&--\\
$B_c\rightarrow D_s^+\bar{K}^0$ &A,P&$2.1^{+0.9+0.3+0.3}_{-0.6-0.3-0.2}$&1.9&--&$54.3^{+6.9+5.3+0.0}_{-7.2-8.0-0.3}$&13.3\\
$B_c\rightarrow D_s^+\eta$ &A,P&$17.3^{+1.7+0.5+3.3}_{-1.8-0.6-1.2}$&--&0.009&$2.8^{+0.0+0.4+1.1}_{-0.1-0.7-1.2}$&--\\
$B_c\rightarrow D_s^+\eta'$ &A,P&$51.0^{+4.9+0.4+6.7}_{-5.4-0.3-3.5}$&--&0.0048&$1.1^{+0.1+0.2+0.7}_{-0.0-0.2-0.6}$&--\\
\bottomrule[2pt]
\end{tabular}
\footnotetext[1]{
we use the results of decay widths in \cite{prd73054024} , but we take $\tau_{B_c}=0.453\text{ps}$
to estimate the branching ratio}
\end{table}
\begin{table}
\caption{ CP averaged branching ratios  and direct CP asymmetries
for $ B_c\rightarrow D_{(s)}V$ decays, together with results from RCQM and LFQM.} \label{tab:bcdv}
\begin{tabular}[t]{l|c|c|c|c|c|c}
\toprule[2pt]
\multicolumn{2}{c|}{}&\multicolumn{3}{c}{ $\mathcal {BR}(10^{-7})$} &\multicolumn{2}{|c}{ $A^{dir}_{CP}(\%)$}  \\ \hline
channels & Class & This work & RCQM & LFQM & This work & RCQM \\ \hline
$B_c\rightarrow D^0\rho^+$ &T &$66.2^{+7.6+16.0+1.6}_{-7.6-14.1-1.3}$ &60.0&13 &$-24.5^{+2.6+5.3+0.3}_{-0.4-3.2-0.8}$ &3.8\\
$B_c\rightarrow D^+\rho^0$ & C,A&$1.4^{+0.1+0.5+0.1}_{-0.2-0.5-0.2}$&3.9&0.2&$79.8^{+0.3+11.2+3.4}_{-5.8-19.6-10.7}$&-3.0\\
$B_c\rightarrow D^0K^{*+}$ &A,P &$25.9^{+2.7+0.9+1.5}_{-3.0-0.8-0.8}$&34.7&0.68&$-66.2^{+1.8+15.1+0.7}_{-0.6-6.5-0.0}$&-6.2\\
$B_c\rightarrow D^+K^{*0}$ & A,P&$19.1^{+3.3+0.1+0.7}_{-2.5-0.0-0.7}$&28.8&--&$3.5^{+0.0+0.5+0.5}_{-0.1-0.8-0.3}$&-0.8\\
$B_c\rightarrow D^+\omega$ &C,A &$1.9^{+0.3+0.5+0.0}_{-0.3-0.6-0.0}$&--&0.15&$-3.6^{+3.9+1.3+13.4}_{-1.2-1.6-10.7}$&--\\
$B_c\rightarrow D^+\phi$   &P &$0.008^{+0.001+0.0+0.001}_{-0.001-0.0-0.001}$&--&--&--&--\\
$B_c\rightarrow D_s^+\rho^0$ &C,P&$0.95^{+0.10+0.02+0.04}_{-0.09-0.01-0.04}$&--&--&$50.2^{+1.0+5.9+2.5}_{-1.1-11.9-3.2}$&--\\
$B_c\rightarrow D_s^+\bar{K}^{*0}$ &A,P&$1.4^{+0.2+0.0+0.1}_{-0.2-0.1-0.1}$&1.0&--&$61.0^{+0.0+6.5+4.5}_{-0.3-14.2-3.6}$&13.3\\
$B_c\rightarrow D_s^+\omega$ &C,P&$0.31^{+0.03+0.07+0.07}_{-0.03-0.07-0.05}$&--&0.016&$44.9^{+0.8+17.1+10.3}_{-1.6-14.9-13.6}$&--\\
$B_c\rightarrow D_s^+\phi$ &A,P&$27.0^{+4.8+0.1+2.0}_{-1.2-0.0-0.4}$&15.7&0.0048&$3.3^{+0.0+0.4+0.3}_{-0.3-0.8-0.4}$&-0.8\\
\bottomrule[2pt]
\end{tabular}
\end{table}

We also cite theoretical
results for
 the   relevant  decays evaluated
 in LFQM model \cite{09095028} and RCQM  model \cite{prd73054024}
 to make a comparison in Tables \ref{tab:bcdp} and \ref{tab:bcdv}.
Our pQCD results are generally close
 to RCQM results but differ substantially from the ones obtained by LFQM.
 This is due to the fact that LFQM used a smaller form factors
 $F^{B_c\rightarrow D} (q^2=0)=0.086$ at maximum recoil, which is rather smaller than
other  model predictions \cite{prd80054016} and also another covariant LFQM results \cite{wang}. In fact, these model calculations all give consistent form factors at the zero coil region, considering only soft contributions by  the
overlap between the initial and final state meson wave functions, which is good  at the zero recoil region.  At the 
maximum-recoil region, which is the case for non-leptonic B decays,  the soft contribution is suppressed, since a hard gluon is needed, as discussed in the previous section.   Furthermore
  LFQM only consider the contribution of current-current operators at the tree
  level, therefore
 they cannot give  predictions for  those modes without tree   diagram  contributions
 like $B_c\rightarrow D^+ K^0$ and   $B_c\rightarrow D_s^+ \bar{K}^0$.
 For the  color suppressed decays (C),
 our predictions  differ from the ones of RCQM, since in these
 modes,
  the contributions from the non-factorizable emission diagram and
  annihilation diagram dominated   the branching ratio,
 which are not calculable in RCQM.

Our numerical results of the CP averaged branching ratios  and direct CP
asymmetries for  $ B_c\rightarrow D^*_{(s)}P$ decays
 are listed
 in Table
\ref{tab:bcdsp}, together with the RCQM model predictions. Again,
our results are similar with RCQM model for the tree dominant mode (T).
But for the annihilation dominant and  penguin dominant modes (A,P), the branching ratios obtained in
the RCQM are one order of magnitude smaller than ours. The reason is
that these decay   amplitudes are governed by the QCD
  penguin parameters $a_4$ and $a_6$ in the combination $a_4+R a_6$ \cite{prd60094014} in the factorization hypothesis.
   The coefficient $R$ arises from the penguin operator
  $O_6$, where  $R>0$ for $B\rightarrow PP$, $R=0$ for $PV$ and $VV$ final states, and $R<0$ for $B\rightarrow VP$,
   the second meson in the final states is the one emitted from vacuum.  Therefore, the branching ratios of various type decays
have the following pattern in the factorization approach
\begin{eqnarray}
\mathcal {BR}(B_c\rightarrow DP)>\mathcal {BR}(B_c\rightarrow
DV)\sim \mathcal {BR}(B_c\rightarrow D^*V)>\mathcal
{BR}(B_c\rightarrow D^*P)\label{penguinrel}
\end{eqnarray}
as a consequence of the interference between the $a_4$ and $a_6$
penguin terms. In the contrary, we have additional non-factorization
contributions and large annihilation type contributions in the pQCD
approach, which spoils the relation in eq.(\ref{penguinrel}).

As expected, the annihilation type diagrams  give large
contributions  in the  $B_c$ meson  decays, because   the
annihilation type diagram contributions  are enhanced by  the CKM
factor $V^*_{cb}V_{cq}$ \cite{prd81014022,plb525240}. For the $b\rightarrow d$
process, $|\frac{V^*_{cb}V_{cd}}{V^*_{ub}V_{ud}}|=2.5$; For the
$b\rightarrow s$ process,
$|\frac{V^*_{cb}V_{cs}}{V^*_{ub}V_{us}}|=47$.
 The annihilation diagram contributions  are the dominant contribution in some $b\rightarrow s$ processes.
Therefore,  we have the ratio relation $\frac{\mathcal
{BR}(B_c\rightarrow D^{(*)0} K^{(*)+})} {\mathcal
{BR}(B_c\rightarrow D^{(*)+} K^{(*)0})}\approx 1$ for these two annihilation dominant
$b\rightarrow s$ transition processes.
\begin{table}
\caption{ CP averaged branching ratios  and direct CP asymmetries
for $ B_c\rightarrow D^*_{(s)}P$ decays, together with results from RCQM. } \label{tab:bcdsp}
\begin{tabular}[t]{l|c|c|c|c|c}
\toprule[2pt]
\multicolumn{2}{c|}{}&\multicolumn{2}{c}{ $\mathcal {BR}(10^{-7})$} &\multicolumn{2}{|c}{ $A^{dir}_{CP}(\%)$}  \\ \hline
channels &Class & This work & RCQM  & This work & RCQM \\ \hline
$B_c\rightarrow D^{*0}\pi^+$ &T& $18.8^{+2.0+4.1+0.4}_{-2.0-3.5-0.5}$ &19.6&$64.0^{+12.0+6.1+0.7}_{-7.6-13.0-0.5}$ &1.5\\
$B_c\rightarrow D^{*+}\pi^0$ &C,A& $1.3^{+0.4+0.2+0.0}_{-0.3-0.3-0.0}$&0.66&$9.6^{+3.3+3.4+10.8}_{-2.7-3.3-8.8}$&-2.1\\
$B_c\rightarrow D^{*0}K^+$ & A,P&$73.5^{+31.0+0.8+0.7}_{-23.4-1.1-0.4}$&4.9&$25.0^{+4.4+3.2+0.1}_{-4.1-6.1-0.3}$&-8.2\\
$B_c\rightarrow D^{*+}K^0$ & A,P&$77.8^{+25.4+0.2+7.2}_{-24.0-0.2-5.2}$&2.8&$-0.3^{+0.0+0.0+0.0}_{-0.0-0.0-0.0}$&-8.2\\
$B_c\rightarrow D^{*+}\eta$ &C,A&$0.34^{+0.14+0.19+0.04}_{-0.09-0.15-0.00}$&--&$-2.0^{+0.0+0.7+22.8}_{-2.4-1.5-30.0}$&--\\
$B_c\rightarrow D^{*+}\eta'$ &C,A&$0.15^{+0.08+0.08+0.03}_{-0.05-0.06-0.01}$&--&$-41.8^{+17.5+13.0+24.3}_{-24.5-13.0-19.2}$&--\\
$B_c\rightarrow D_s^{*+}\pi^0$ &C,P&$0.27^{+0.02+0.03+0.01}_{-0.04-0.02-0.02}$&--&$29.9^{+2.4+5.3+1.8}_{-1.9-8.2-1.5}$&--\\
$B_c\rightarrow D_s^{*+}\bar{K}^0$&A,P &$1.6^{+0.2+0.1+0.2}_{-0.3-0.1-0.1}$&0.21&$-3.3^{+0.4+0.6+0.9}_{-1.0-0.4-0.4}$&13.3\\
$B_c\rightarrow D_s^{*+}\eta$ &A,P&$16.7^{+5.3+0.3+0.1}_{-4.0-0.2-0.3}$&--&$-0.7^{+0.2+0.2+0.6}_{-0.2-0.0-0.3}$&--\\
$B_c\rightarrow D_s^{*+}\eta'$ &A,P&$14.4^{+6.6+0.1+0.5}_{-4.6-0.1-0.6}$&--&$0.02^{+0.01+0.00+0.55}_{-0.02-0.01-0.52}$&--\\
\bottomrule[2pt]
\end{tabular}
\end{table}

For the  color suppressed decays (C),
 our predictions  differ from the ones of RCQM, since in these
 modes,
  the contributions from the non-factorizable emission diagram and
  annihilation diagram dominated   the branching ratio,
 which are not calculable in RCQM.
  For example in decays $B_c\rightarrow D^{(*)+} (\pi^0, \eta, \eta', \rho^0, \omega)$, the non-factorizable contribution, which is proportional
to the large Wilson coefficient $C_2$, is the dominant contribution.
In fact,
the annihilation diagrams can also give relatively large
contributions for the enhancement by CKM factor.
We also find that the twist-3 distribution amplitudes play
an important role, especially in the factorizable  annihilation
diagrams.
 As stated  in section \ref{sec:bcdsp}, the $B_c\rightarrow DP(V)$ decay amplitudes are different from $B_c\rightarrow D^*P(V)$
 ones only at twist-3 level.
 The numerical results show that
 the contributions from factorizable annihilation diagrams have an opposite sign between the two type channels. For example, this results
 in a constructive interference between non-factorizable  emission diagrams and factorizable annihilation diagrams for $B_c\rightarrow D^{*+}\pi^0$,
 but a destructive interference for $B_c\rightarrow D^{+}\pi^0$.
This makes $\mathcal {BR}(B_c\rightarrow D^{*+}\pi^0)$
 larger than $\mathcal {BR}(B_c\rightarrow D^{+}\pi^0)$.  Similarly, we have
 $\mathcal {BR}(B_c\rightarrow D^{*+}\rho^0)> \mathcal {BR}(B_c\rightarrow D^{+}\rho^0)$.
  However, for $B_c\rightarrow D^{(*)+}\eta(\eta')$,
while the $d\bar{d}$ part contributes to the annihilation diagrams, the constructive or destructive interference situation are just the reverse,
and  $\mathcal {BR}(B_c\rightarrow D^{*+}\eta(\eta'))$ are smaller than $\mathcal {BR}(B_c\rightarrow D^{+}\eta(\eta'))$.

\begin{table}
\caption{ CP averaged branching ratios, direct CP asymmetries   and
the transverse polarizations fractions for $ B_c\rightarrow
D^*_{(s)}V$ decays, together with results from RCQM.} \label{tab:bcdsv}
\begin{tabular}[t]{l|c|c|c|c|c|c}
\toprule[2pt] \multicolumn{2}{c|}{}&\multicolumn{2}{c|}{ $\mathcal {BR}(10^{-7})$} &
\multicolumn{2}{|c|}{ $A^{dir}_{CP}(\%)$}  &$\mathcal {R}_T(\%)$
\\ \hline channels &Class& This work & RCQM   & This work & RCQM  & This
work \\ \hline
$B_c\rightarrow D^{*0}\rho^+$ &T& $55.3^{+8.6+11.9+1.5}_{-5.0-11.1-1.4}$ &59.7&$-24.1^{+3.0+4.2+0.4}_{-3.4-2.7-0.4}$ & 3.8&$16.4^{+2.5+2.0+0.3}_{-1.7-1.4-0.1}$\\
$B_c\rightarrow D^{*+}\rho^0$ &C,A& $3.8^{+1.0+0.5+0.1}_{-0.8-0.6-0.1}$&13.0&$30.2^{+0.0+2.6+5.4}_{-1.5-5.8-7.6}$& -3.0&$54.3^{+1.8+4.0+0.5}_{-0.9-2.4-0.4}$\\
$B_c\rightarrow D^{*0}K^{*+}$ &A,P& $161^{+59+5+11}_{-40-4-9}$&37.7&$-14.9^{+1.1+3.1+0.3}_{-0.8-1.7-0.1}$& -6.2&$52.6^{+1.5+2.3+1.3}_{-1.1-1.8-0.7}$\\
$B_c\rightarrow D^{*+}K^{*0}$ &A,P& $172^{+57+1+11}_{-42-1-9}$&30.6&$0.4^{+0.0+0.0+0.0}_{-0.0-0.1-0.0}$ &-0.8 &$57.4^{+0.6+0.1+0.9}_{-0.7-0.1-0.4}$\\
$B_c\rightarrow D^{*+}\omega$ &C,A&$2.4^{+0.4+0.9+0.2}_{-0.6-0.7-0.1}$&--&$ -7.8^{+1.0+2.6+5.8}_{-0.0-3.4-5.0}$&--&$56.0^{+1.2+9.6+0.7}_{-0.5-6.5-0.7}$\\
$B_c\rightarrow D^{*+}\phi$   &P&$0.004^{+0.001+0+0}_{-0-0-0.001}$&--&-- &-- &$11.4^{+22.3+0.0+5.3}_{-5.4-0.0-6.9}$\\
$B_c\rightarrow D_s^{*+}\rho^0$ &C,P&$0.72^{+0.08+0.03+0.02}_{-0.08-0.03-0.03}$&--& $-29.3^{+1.3+7.6+1.4}_{-1.1-4.5-0.9}$&-3.0 &$11.2^{+0.5+2.1+0.2}_{-0.3-1.4-0.1}$\\
$B_c\rightarrow D_s^{*+}\bar{K}^{*0}$ &A,P&$4.3^{+1.3+0.4+0.3}_{-1.0-0.3-0.2}$&2.9&$6.2^{+0.1+1.3+0.0}_{-0.3-1.8-0.1}$& 13.3&$68.8^{+2.1+3.9+0.8}_{-2.3-4.4-0.4}$\\
$B_c\rightarrow D_s^{*+}\omega$ &C,P&$0.26^{+0.03+0.04+0.07}_{-0.01-0.05-0.04}$&--&$-21.3^{+5.3+6.8+7.9}_{-4.6-6.8-4.7}$&--&$49.5^{+8.8+2.1+4.4}_{-10.9-1.2-2.8}$\\
$B_c\rightarrow D_s^{*+}\phi$ &A,P&$137.3^{+39.3+0.5+10.5}_{-27.8-0.5-7.5}$&38.8&$0.3^{+0.1+0.1+0.0}_{-0.1-0.1-0.0}$ & -0.8&$67.5^{+2.1+0.1+1.4}_{-3.1-0.2-1.5}$\\
\bottomrule[2pt]
\end{tabular}
\end{table}

 For another kind of $b\rightarrow s$ processes, the decays   $B_c\rightarrow
D_s^{(*)+}(\pi^0,\rho^0,\omega)$ have a small branching ratio at
$\mathcal {O}(10^{-8})$ due to the absent  annihilation diagram
contributions, and  the emission diagram contributions   suppressed
by CKM matrix elements $|V^*_{ub}V_{us}|$. Since the contribution of
penguin operator is comparable to the one of tree operator,  the
interference between the two contributions is large. As a result, a
big CP asymmetry is predicted in these decays.
The branching ratio is  even smaller $\sim 10^{-10}$ and no CP
violation for $B_c\rightarrow D^{(*)+}\phi$ decays, since there are
only penguin diagrams contributions. All these and other rare decays
are also important, since they are quite sensitive to new physics
contributions.

For the $B_c\rightarrow D^*_{(s)}V$ decays the branching ratios  and
the transverse polarization fractions $\mathcal {R}_T$ are given as
\begin{eqnarray}
\mathcal {BR}=\frac{G_F\tau_{B_c}}{32\pi
M_B}(1-r_D^2)\sum_{i=0,+,-}|\mathcal {A}_i|^2,\quad \mathcal
{R}_T=\frac{|\mathcal {A}_+|^2+|\mathcal {A}_-|^2}{|\mathcal
{A}_0|^2+|\mathcal {A}_+|^2+|\mathcal {A}_-|^2},
\end{eqnarray}
where  the helicity amplitudes $\mathcal {A}_i$  have the following
relationships with $\mathcal {A}^{L,N,T}$
\begin{eqnarray}
\mathcal {A}_0=\mathcal {A}^L,\quad \mathcal {A}_{\pm}=\mathcal
{A}^N \pm \mathcal {A}^T .
\end{eqnarray}

 Our numerical results of the CP averaged branching ratios, direct CP asymmetries  and
the transverse polarization fractions for $ B_c\rightarrow
D^*_{(s)}V$ decays are shown in Table~\ref{tab:bcdsv}.   The
transverse polarization contributions  are usually suppressed by the
factor $r_V$ or $r_{D^*}$ comparing with the longitudinal polarization
contributions,  thus
  we do have relatively small
transverse polarization factions for the tree-dominant decay
($\mathcal {R}_T(B_c\rightarrow D^{*0} \rho^+)=16.4\%$) and the pure
penguin type decay ($\mathcal {R}_T(B_c\rightarrow D^{*+}
\phi)=11.5\%$). For the pure emission type decay  $B_c\rightarrow
D^{*+}_s \omega $, the  transverse polarization faction is   large
because the non-factorizable emission diagram induced by operate
$O_6$ can enhance the transverse polarization sizably. The fact that
the non-factorizable contribution can give large transverse
polarization contribution is also observed in the $B^0 \to
\rho^0\rho^0$, $\omega\omega$ decays \cite{rhorho}.
 For other   decays,
the   annihilation type contributions  dominate the branching ratios
due to the large Wilson coefficients.    Therefore, the transverse
polarizations take a  larger ratio in the branching ratios, which
can reach $50\%\sim 70\%$. This is similar to the case of $B\to \phi
K^*$ and various $B \to \rho K^*$ decays \cite{prd66054013,0508080}.

 From Table \ref{tab:bcdsv}, one can also see that our branching ratios for   $B_c\rightarrow
D^{*+}K^{*0},D^{*0}K^{*+},D_s^{*+}\bar{K}^{*0},D^{*+}_s\phi$ decays,
 are about 2 to 5 times larger than those in
RCQM model, due to the sizable contributions of transverse polarization
amplitudes. Another point should be addressed that the annihilation
contributions with a strong phase have remarkable effects on the
direct CP asymmetries in these decays. As a result, our predictions
are somewhat larger than those  from RCQM.

\section{ conclusion}

In this paper, we  investigate the two body non-leptonic decays of
$B_c$ meson with the final states involving one $D_{(s)}^{(*)}$
meson in the pQCD approach based on $k_T$ factorization. It is found
that the non-factorizable emission and annihilation type diagrams are
possible to give a large contribution, especially for those color
suppressed modes and annihilation diagram dominant modes. All the
branching ratios and CP asymmetry parameters are calculated and the
ratios of the transverse polarization contributions in the
$B_c\rightarrow D_{(s)}^*V$ decays are  estimated.  Because of the different
weak phase and strong phase from tree diagrams, penguin diagrams and
annihilation diagrams, we predict a possible large direct
CP-violation in some channels.   We also find that the transverse
polarization contributions in some channels, which mainly come from the
non-factorizable emission diagrams or annihilation type diagrams,
are large.

Generally, our predictions for the branching
ratios in the tree-dominant $B_c$ decays are in good agreements
with that of RCQM model. But we have much larger branching ratios
  in the color-suppressed,
annihilation diagram dominant $B_c$ decays, due to the included
non-factorizable diagrams and annihilation type diagrams
contributions.

\begin{acknowledgments}
We thank  Hsiang-nan Li and Fusheng Yu for helpful
discussions. This work is partially supported by
 National Natural
Science Foundation of China under the Grant No. 10735080, and
11075168; Natural Science Foundation of Zhejiang Province of China,
  Grant No. Y606252 and Scientific Research Fund of Zhejiang Provincial Education Department of China, Grant No. 20051357.
\end{acknowledgments}
\begin{appendix}
\section{factorization formulas for $B_c\rightarrow D^*V$}\label{sec:a}
We mark L, N and T to denote the contributions from longitudinal polarization, normal polarization and transverse polarization, respectively.
\begin{eqnarray}\label{eq:flll}
\mathcal {F}^{LL,L}_e&=&2\sqrt{\frac{2}{3}}\pi C_f f_Bf_VM^4_B\int_0^1dx_2\int_0^{\infty}b_1b_2db_1db_2\phi_{D^*_{(s)}}(x_2,b_2)\times\nonumber\\&&\{[
x_2-2r_b+r_D(r_b-2x_2)]\alpha_s(t_a)h_e(\alpha_e,\beta_a,b_1,b_2)S_t(x_2)\exp[-S_{ab}(t_a)]\nonumber\\&&
+[r_D^2(x_1-1)]\alpha_s(t_b)h_e(\alpha_e,\beta_b,b_2,b_1)S_t(x_1)\exp[-S_{ab}(t_b)]\},
\end{eqnarray}
\begin{eqnarray}
\mathcal {F}^{LL,N}_e&=&2\sqrt{\frac{2}{3}}\pi C_f f_Bf_VM^4_Br_V\int_0^1dx_2\int_0^{\infty}b_1b_2db_1db_2\phi_{D^*_{(s)}}(x_2,b_2)\times\nonumber\\&&\{[
r_b-2+r_D(x_2+1-4r_b)]\alpha_s(t_a)h_e(\alpha_e,\beta_a,b_1,b_2)S_t(x_2)\exp[-S_{ab}(t_a)\nonumber\\&&
+r_D[2x_1-1]\alpha_s(t_b)h_e(\alpha_e,\beta_b,b_2,b_1)S_t(x_1)\exp[-S_{ab}(t_b)]\},
\end{eqnarray}
\begin{eqnarray}
\mathcal {F}^{LL,T}_e&=&2\sqrt{\frac{2}{3}}\pi C_f f_Bf_VM^4_Br_V\int_0^1dx_2\int_0^{\infty}b_1b_2db_1db_2\phi_{D^*_{(s)}}(x_2,b_2)\times\nonumber\\&&\{[
r_b-2+r_D(1-x_2)]\alpha_s(t_a)h_e(\alpha_e,\beta_a,b_1,b_2)S_t(x_2)\exp[-S_{ab}(t_a)]\nonumber\\&&
-r_D\alpha_s(t_b)h_e(\alpha_e,\beta_b,b_2,b_1)S_t(x_1)\exp[-S_{ab}(t_b)]\},
\end{eqnarray}
\begin{eqnarray}
\mathcal {F}^{LR,L}_e&=&\mathcal {F}^{LL,L}_e,\quad \mathcal {F}^{LR,N}_e=\mathcal {F}^{LL,N}_e,\quad \mathcal {F}^{LR,T}_e=\mathcal {F}^{LL,T}_e.
\end{eqnarray}
The factorizable emission topology contribution $\mathcal {F}^{SP,i}_e(i=L,N,T)$ vanishes due to the conservation of charge parity.
\begin{eqnarray}
\mathcal {M}^{LL,L}_e&=&-\frac{8}{3}\pi C_f f_BM^4_B\int_0^1dx_2dx_3\int_0^{\infty}b_2b_3db_2db_3\phi_{D^*_{(s)}}(x_2,b_2)\phi_V(x_3)\{[1-x_1\nonumber\\&&
-x_3-r_D(x_1+x_2-1)]\alpha_s(t_c)h_e(\beta_c,\alpha_e,b_3,b_2)\exp[-S_{cd}(t_c)]+
[-1+2x_1\nonumber\\&&+x_2-x_3-r_D(x_1+x_2-1)]\alpha_s(t_d)h_e(\beta_d,\alpha_e,b_3,b_2)\exp[-S_{cd}(t_d)]\},
\end{eqnarray}
\begin{eqnarray}
\mathcal {M}^{LL,N}_e&=&\frac{8}{3}\pi C_f f_BM^4_Br_V\int_0^1dx_2dx_3\int_0^{\infty}b_2b_3db_2db_3\phi_{D^*_{(s)}}(x_2,b_2)\{[
(x_1+x_3-1)\nonumber\\&&\phi_V^{\nu}(x_3)+2r_D(x_3-x_2)\phi_V^a(x_3)]\alpha_s(t_c)h_e(\beta_c,\alpha_e,b_3,b_2)\exp[-S_{cd}(t_c)]\nonumber\\&&
-[(-2r_D(1-2x_1-x_2+x_3)-x_1+x_3)\phi_V^{\nu}(x_3)+2(r_D(1-x_2-x_3)\nonumber\\&&-2x_1+2x_3)\phi_V^a(x_3)]
\alpha_s(t_d)h_e(\beta_d,\alpha_e,b_3,b_2)\exp[-S_{cd}(t_d)]\},
\end{eqnarray}
\begin{eqnarray}
\mathcal {M}^{LL,T}_e&=&\frac{8}{3}\pi C_f f_BM^4_Br_V\int_0^1dx_2dx_3\int_0^{\infty}b_2b_3db_2db_3\phi_{D^*_{(s)}}(x_2,b_2)\times\nonumber\\&&\{[
(x_1+x_3-1)\phi_V^{\nu}(x_3)-2r_D(2x_1+x_2+x_3-2)\phi_V^a(x_3)]\nonumber\\&&\alpha_s(t_c)h_e(\beta_c,\alpha_e,b_3,b_2)\exp[-S_{cd}(t_c)]\nonumber\\&&
-[(x_3-x_1)\phi_V^{\nu}(x_3)+2(r_D(2x_1+x_2-x_3-1)-2x_1+2x_3)\phi_V^a(x_3)]\nonumber\\&&
\alpha_s(t_d)h_e(\beta_d,\alpha_e,b_3,b_2)\exp[-S_{cd}(t_d)]\},
\end{eqnarray}
\begin{eqnarray}
\mathcal {M}^{LR,L}_e&=&-\frac{8}{3}\pi C_f f_BM^4_B\int_0^1dx_2dx_3\int_0^{\infty}b_2b_3db_2db_3\phi_{D^*_{(s)}}(x_2,b_2)\phi_V(x_3)\times\nonumber\\&&\{[
(x_1+x_3-1+r_D(x_2-x_3))\phi_V^s(x_3)+(x_1+x_3-1-\nonumber\\&&r_D(2x_1+x_2+x_3-2))\phi_V^t(x_3)]
\alpha_s(t_c)h_e(\beta_c,\alpha_e,b_3,b_2)\exp[-S_{cd}(t_c)]\nonumber\\&&
-[(x_1-x_3-r_D(1-x_2-x_3))\phi_V^s(x_3)-(x_1-x_3+r_D(1-2x_1\nonumber\\&&-x_2+x_3))\phi_V^t(x_3)]
\alpha_s(t_d)h_e(\beta_d,\alpha_e,b_3,b_2)\exp[-S_{cd}(t_d)]\},
\end{eqnarray}
\begin{eqnarray}
\mathcal {M}^{LR,T}_e&=&\mathcal {M}^{LR,N}_e=-\frac{8}{3}\pi C_f f_BM^4_Br_D\int_0^1dx_2dx_3\int_0^{\infty}b_2b_3db_2db_3
\phi_{D^*_{(s)}}(x_2,b_2)\nonumber\\&&\phi_V^T(x_3)(x_1+x_2-1)\{
\alpha_s(t_c)h_e(\beta_c,\alpha_e,b_3,b_2)\exp[-S_{cd}(t_c)]\nonumber\\&&+
\alpha_s(t_d)h_e(\beta_d,\alpha_e,b_3,b_2)\exp[-S_{cd}(t_d)]\},
\end{eqnarray}
\begin{eqnarray}
\mathcal {M}^{SP,L}_e&=&-\frac{8}{3}\pi C_f f_BM^4_B\int_0^1dx_2dx_3\int_0^{\infty}b_2b_3db_2db_3\phi_{D^*_{(s)}}(x_2,b_2)\phi_V(x_3)\{[
2-2x_1\nonumber\\&&-x_2-x_3+r_D(x_1+x_2-1)]
\alpha_s(t_c)h_e(\beta_c,\alpha_e,b_3,b_2)\exp[-S_{cd}(t_c)]-\nonumber\\&&
[x_3-x_1-r_D(x_1+x_2-1)]
\alpha_s(t_d)h_e(\beta_d,\alpha_e,b_3,b_2)\exp[-S_{cd}(t_d)]\},
\end{eqnarray}
\begin{eqnarray}
\mathcal {M}^{SP,N}_e&=&-\frac{8}{3}\pi C_f f_BM^4_Br_V\int_0^1dx_2dx_3\int_0^{\infty}b_2b_3db_2db_3
\phi_{D^*_{(s)}}(x_2,b_2)\times\nonumber\\&&\{[(x_1+x_3-1-2r_D(2x_1+x_2+x_3-2))
\phi_V^{\nu}(x_3)-(x_1+x_3-1)\phi_V^a(x_3)]\nonumber\\&&
\alpha_s(t_c)h_e(\beta_c,\alpha_e,b_3,b_2)\exp[-S_{cd}(t_c)]\nonumber\\&&+(x_1-x_3)(\phi_V^{\nu}(x_3)-\phi_V^{a}(x_3))
\alpha_s(t_d)h_e(\beta_d,\alpha_e,b_3,b_2)\exp[-S_{cd}(t_d)]\},
\end{eqnarray}
\begin{eqnarray}
\mathcal {M}^{SP,T}_e&=&\frac{8}{3}\pi C_f f_BM^4_Br_V\int_0^1dx_2dx_3\int_0^{\infty}b_2b_3db_2db_3
\phi_{D^*_{(s)}}(x_2,b_2)\times\nonumber\\&&\{[(x_1+x_3-1-2r_D(2x_1+x_2+x_3-2))
\phi_V^a(x_3)-(x_1+x_3-1)\phi_V^{\nu}(x_3)]\nonumber\\&&
\alpha_s(t_c)h_e(\beta_c,\alpha_e,b_3,b_2)\exp[-S_{cd}(t_c)]\nonumber\\&&+(x_1-x_3)(\phi_V^{a}(x_3)-\phi_V^{\nu}(x_3))
\alpha_s(t_d)h_e(\beta_d,\alpha_e,b_3,b_2)\exp[-S_{cd}(t_d)]\},
\end{eqnarray}

\begin{eqnarray}
\mathcal {F}_a^{LL,L}&=&8C_Ff_B\pi M_B^4
\int_0^1dx_2dx_3\int_0^{\infty}b_2b_3db_2db_3\phi_{D^*_{(s)}}(x_2,b_2)
\{[-x_3\phi_V(x_3)\nonumber\\&&+r_cr_V(\phi_V^t(x_3)-\phi_V^s(x_3))]\alpha_s(t_e)h_e(\alpha_a,\beta_e,b_2,b_3)\exp[-S_{ef}(t_e)]S_t(x_3)+
\nonumber\\&&[x_2\phi_V(x_3)+2r_Vr_D(x_2-1)\phi_V^s(x_3)]\nonumber\\&&\alpha_s(t_f)h_e(\alpha_a,\beta_f,b_3,b_2)\exp[-S_{ef}(t_f)]S_t(x_2)
\},
\end{eqnarray}

\begin{eqnarray}
\mathcal {F}_a^{LL,N}&=&-8C_Ff_B\pi M_B^4r_D
\int_0^1dx_2dx_3\int_0^{\infty}b_2b_3db_2db_3\phi_{D^*_{(s)}}(x_2,b_2)
\{[r_V(x_3-1)\phi_V^a(x_3)\nonumber\\&&-r_c\phi_V^T(x_3)+r_V(x_3+1)\phi_V^{\nu}]\alpha_s(t_e)h_e(\alpha_a,\beta_e,b_2,b_3)\exp[-S_{ef}(t_e)]S_t(x_3)
\nonumber\\&&-r_V[(x_2+1)\phi_V^{\nu}(x_3)-(x_2-1)\phi_V^a(x_3)]\nonumber\\&&\alpha_s(t_f)h_e(\alpha_a,\beta_f,b_3,b_2)\exp[-S_{ef}(t_f)]S_t(x_2)
\},
\end{eqnarray}
\begin{eqnarray}
\mathcal {F}_a^{LL,T}&=&8C_Ff_B\pi M_B^4r_D
\int_0^1dx_2dx_3\int_0^{\infty}b_2b_3db_2db_3\phi_{D^*_{(s)}}(x_2,b_2)\times\nonumber\\&&
\{[r_V(x_3+1)\phi_V^a(x_3)-r_c\phi_V^T(x_3)+r_V(x_3-1)\phi_V^{\nu}(x_3)]\nonumber\\&&\alpha_s(t_e)h_e(\alpha_a,\beta_e,b_2,b_3)\exp[-S_{ef}(t_e)]S_t(x_3)
\nonumber\\&&+r_V[(-x_2-1)\phi_V^a(x_3)+(x_2-1)\phi_V^{\nu}(x_3)]\nonumber\\&&\alpha_s(t_f)h_e(\alpha_a,\beta_f,b_3,b_2)\exp[-S_{ef}(t_f)]S_t(x_2)
\},
\end{eqnarray}
\begin{eqnarray}
\mathcal {F}_a^{LR,L}&=&\mathcal {F}_a^{LL,L},\quad \mathcal {F}_a^{LR,N}=\mathcal {F}_a^{LL,N},\quad \mathcal {F}_a^{LR,T}=\mathcal {F}_a^{LL,T},
\end{eqnarray}
\begin{eqnarray}
\mathcal {F}_a^{SP,L}&=&16C_Ff_B\pi M_B^4
\int_0^1dx_2dx_3\int_0^{\infty}b_2b_3db_2db_3\phi_{D^*_{(s)}}(x_2,b_2)
\{[r_c\phi_V(x_3)\nonumber\\&&+r_Vx_3(\phi_V^s(x_3)-\phi_V^t(x_3))]\alpha_s(t_e)h_e(\alpha_a,\beta_e,b_2,b_3)\exp[-S_{ef}(t_e)]-
\nonumber\\&&[r_Dx_2\phi_V(x_3)-2r_V\phi_V^s(x_3)]\alpha_s(t_f)h_e(\alpha_a,\beta_f,b_3,b_2)\exp[-S_{ef}(t_f)]
\},
\end{eqnarray}
\begin{eqnarray}
\mathcal {F}_a^{SP,N}&=&-16C_Ff_B\pi M_B^4
\int_0^1dx_2dx_3\int_0^{\infty}b_2b_3db_2db_3\phi_{D^*_{(s)}}(x_2,b_2)\times\nonumber\\&&
\{r_D[\phi_V^T(x_3)-2r_cr_V\phi_V^{\nu}(x_3)]\alpha_s(t_e)h_e(\alpha_a,\beta_e,b_2,b_3)\exp[-S_{ef}(t_e)]
\nonumber\\&&+r_V(\phi_V^{\nu}(x_3)+\phi_V^a(x_3))\alpha_s(t_f)h_e(\alpha_a,\beta_f,b_3,b_2)\exp[-S_{ef}(t_f)]
\},
\end{eqnarray}
\begin{eqnarray}
\mathcal {F}_a^{SP,T}&=&-16C_Ff_B\pi M_B^4
\int_0^1dx_2dx_3\int_0^{\infty}b_2b_3db_2db_3\phi_{D^*_{(s)}}(x_2,b_2)\times\nonumber\\&&
\{r_D[\phi_V^T(x_3)-2r_cr_V\phi_V^{a}(x_3)]\alpha_s(t_e)h_e(\alpha_a,\beta_e,b_2,b_3)\exp[-S_{ef}(t_e)]
\nonumber\\&&+r_V(\phi_V^{\nu}(x_3)+\phi_V^a(x_3))\alpha_s(t_f)h_e(\alpha_a,\beta_f,b_3,b_2)\exp[-S_{ef}(t_f)]
\},
\end{eqnarray}
\begin{eqnarray}
\mathcal {M}_a^{LL,L}&=&\frac{8}{3}C_Ff_B\pi M_B^4
\int_0^1dx_2dx_3\int_0^{\infty}b_1b_2db_1db_2\phi_{D^*_{(s)}}(x_2,b_2)
\{[(x_1-x_2-r_c)\nonumber\\&&\phi_V(x_3)-r_Dr_V[(x_2-x_3)\phi_V^s(x_3)-(2x_1-x_2-x_3)\phi_V^t(x_3)]]
\nonumber\\&&\alpha_s(t_g)h_e(\beta_g,\alpha_a,b_1,b_2)\exp[-S_{gh}(t_g)]-
[(1-r_b-x_1-x_3)\phi_V(x_3)\nonumber\\&&-r_Dr_V[(x_3-x_2)\phi_V^s(x_3)+(2x_1+x_2+x_3-2)\phi_V^t(x_3)]]\nonumber\\&&
\alpha_s(t_h)h_e(\beta_h,\alpha_a,b_1,b_2)\exp[-S_{gh}(t_h)]
\},
\end{eqnarray}
\begin{eqnarray}
\mathcal {M}_a^{LL,N}&=&-\frac{16}{3}C_Ff_B\pi M_B^4r_Dr_V
\int_0^1dx_2dx_3\int_0^{\infty}b_1b_2db_1db_2\phi_{D^*_{(s)}}(x_2,b_2)\phi_V^{\nu}(x_3)\times\nonumber\\&&
\{r_c
\alpha_s(t_g)h_e(\beta_g,\alpha_a,b_1,b_2)\exp[-S_{gh}(t_g)]\nonumber\\&&-
r_b
\alpha_s(t_h)h_e(\beta_h,\alpha_a,b_1,b_2)\exp[-S_{gh}(t_h)]
\},
\end{eqnarray}
\begin{eqnarray}
\mathcal {M}_a^{LL,T}&=&-\frac{16}{3}C_Ff_B\pi M_B^4r_Dr_V
\int_0^1dx_2dx_3\int_0^{\infty}b_1b_2db_1db_2\phi_{D^*_{(s)}}(x_2,b_2)\phi_V^{a}(x_3)\times\nonumber\\&&
\{r_c
\alpha_s(t_g)h_e(\beta_g,\alpha_a,b_1,b_2)\exp[-S_{gh}(t_g)]\nonumber\\&&-
r_b
\alpha_s(t_h)h_e(\beta_h,\alpha_a,b_1,b_2)\exp[-S_{gh}(t_h)]
\},
\end{eqnarray}

\begin{eqnarray}
\mathcal {M}_a^{LR,L}&=&\frac{8}{3}C_Ff_B\pi M_B^4
\int_0^1dx_2dx_3\int_0^{\infty}b_1b_2db_1db_2\phi_{D^*_{(s)}}(x_2,b_2)
\{[r_D(x_1-x_2+r_c)\nonumber\\&&\phi_V(x_3)+r_V(-x_1+x_3-r_c)(\phi_V^s(x_3)+\phi_V^t(x_3))]
\nonumber\\&&\alpha_s(t_g)h_e(\beta_g,\alpha_a,b_1,b_2)\exp[-S_{gh}(t_g)]-
[-r_D(x_1+x_2-r_b-1)\phi_V(x_3)\nonumber\\&&+r_V(x_1+x_3-r_b-1)(\phi_V^s(x_3)+\phi_V^t(x_3))]\nonumber\\&&
\alpha_s(t_h)h_e(\beta_h,\alpha_a,b_1,b_2)\exp[-S_{gh}(t_h)]
\},
\end{eqnarray}
\begin{eqnarray}
\mathcal {M}_a^{LR,T}&=&\mathcal {M}_a^{LR,N}=\frac{8}{3}C_Ff_B\pi M_B^4
\int_0^1dx_2dx_3\int_0^{\infty}b_1b_2db_1db_2\phi_{D^*_{(s)}}(x_2,b_2)\times\nonumber\\&&
\{[r_V(x_1-x_3+r_c)(\phi_V^{\nu}(x_3)+\phi_V^{a}(x_3))-r_D(x_1-x_2+r_c)\phi_V^{T}(x_3)]
\nonumber\\&&\alpha_s(t_g)h_e(\beta_g,\alpha_a,b_1,b_2)\exp[-S_{gh}(t_g)]+\nonumber\\&&
[r_V(x_1+x_3-r_b-1)(\phi_V^{\nu}(x_3)+\phi_V^{a}(x_3))+r_D(1+r_b-x_1-x_2)\phi_V^{T}(x_3)]
\nonumber\\&&\alpha_s(t_h)h_e(\beta_h,\alpha_a,b_1,b_2)\exp[-S_{gh}(t_h)]
\},
\end{eqnarray}
\begin{eqnarray}
\mathcal {M}_a^{SP,L}&=&\frac{8}{3}C_Ff_B\pi M_B^4
\int_0^1dx_2dx_3\int_0^{\infty}b_1b_2db_1db_2\phi_{D^*_{(s)}}(x_2,b_2)
\{[(x_1-x_3-r_c)\nonumber\\&&\phi_V(x_3)-r_Dr_V[(x_3-x_2)\phi_V^s(x_3)-(2x_1-x_2-x_3)\phi_V^t(x_3)]]
\nonumber\\&&\alpha_s(t_g)h_e(\beta_g,\alpha_a,b_1,b_2)\exp[-S_{gh}(t_g)]-
[(1-r_b-x_1-x_2)\phi_V(x_3)\nonumber\\&&-r_Dr_V[(x_2-x_3)\phi_V^s(x_3)+(2x_1+x_2+x_3-2)\phi_V^t(x_3)]]\nonumber\\&&
\alpha_s(t_h)h_e(\beta_h,\alpha_a,b_1,b_2)\exp[-S_{gh}(t_h)]
\},
\end{eqnarray}
\begin{eqnarray}
\mathcal {M}_a^{SP,N}=\mathcal {M}_a^{LL,N},\quad \mathcal {M}_a^{SP,T}=-\mathcal {M}_a^{LL,T}.
\end{eqnarray}

\section{scales and related functions in hard kernel }\label{sec:b}
We show here the functions $h_e$, coming from the Fourier transform of hard kernel.
\begin{eqnarray}
h_e(\alpha,\beta,b_1,b_2)&=&h_1(\alpha,b_1)\times h_2(\beta,b_1,b_2),\nonumber\\
h_1(\alpha,b_1)&=&\left\{\begin{array}{ll}
K_0(\sqrt{\alpha}b_1), & \quad  \quad \alpha >0\\
K_0(i\sqrt{-\alpha}b_1),& \quad  \quad \alpha<0
\end{array} \right.\nonumber\\
h_2(\beta,b_1,b_2)&=&\left\{\begin{array}{ll}
\theta(b_1-b_2)I_0(\sqrt{\beta}b_2)K_0(\sqrt{\beta}b_1)+(b_1\leftrightarrow b_2), & \quad   \beta >0\\
\theta(b_1-b_2)J_0(\sqrt{-\beta}b_2)K_0(i\sqrt{-\beta}b_1)+(b_1\leftrightarrow b_2),& \quad   \beta<0
\end{array} \right.
\end{eqnarray}
where $J_0$ is the Bessel function and $K_0$, $I_0$ are modified Bessel function with
$K_0(ix)=\frac{\pi}{2}(-N_0(x)+i J_0(x))$.
The hard scale t is chosen as the maximum of the virtuality of the internal momentum transition in the hard amplitudes,
including $1/b_i(i=1,2,3)$:
\begin{eqnarray}
t_a&=&\max(\sqrt{|\alpha_e|},\sqrt{|\beta_a|},1/b_1,1/b_2),\quad t_b=\max(\sqrt{|\alpha_e|},\sqrt{|\beta_b|},1/b_1,1/b_2),\nonumber\\
t_c&=&\max(\sqrt{|\alpha_e|},\sqrt{|\beta_c|},1/b_2,1/b_3),\quad t_d=\max(\sqrt{|\alpha_e|},\sqrt{|\beta_d|},1/b_2,1/b_3),\nonumber\\
t_e&=&\max(\sqrt{|\alpha_a|},\sqrt{|\beta_e|},1/b_2,1/b_3),\quad t_f=\max(\sqrt{|\alpha_a|},\sqrt{|\beta_f|},1/b_2,1/b_3),\nonumber\\
t_g&=&\max(\sqrt{|\alpha_a|},\sqrt{|\beta_g|},1/b_1,1/b_2),\quad t_h=\max(\sqrt{|\alpha_a|},\sqrt{|\beta_h|},1/b_1,1/b_2),
\end{eqnarray}
where
\begin{eqnarray}\label{eq:betai}
\alpha_e&=&(1-x_1-x_2)(x_1-r_D^2)M_B^2,\quad \alpha_a=-x_2x_3(1-r_D^2)M_B^2,\nonumber\\
\beta_a&=&[r_b^2-x_2(1-r_D^2)]M_B^2,\quad \beta_b=-(1-x_1)(x_1-r_D^2)M_B^2,\nonumber\\
\beta_c&=&-(1-x_1-x_2)[1-x_1-x_3(1-r_D^2)]M_B^2,\nonumber\\\quad \beta_d&=&(1-x_1-x_2)[x_1-x_3-r_D^2(1-x_3)]M_B^2,\nonumber\\
\beta_e&=&[r_c^2-x_3-(1-x_3)r_D^2]M_B^2,\quad \beta_f=-x_2(1-r_D^2)]M_B^2,\nonumber\\
\beta_g&=&[r_c^2-(x_1-x_3(1-r_D^2))(x_1-x_2)]M_B^2,\quad \nonumber\\\beta_h&=&[r_b^2-(1-x_1-x_3+x_3r_D^2)(1-x_1-x_2)]M_B^2.
\end{eqnarray}
The Sudakov factors used in the text are defined by
\begin{eqnarray}
S_{ab}(t)&=&s(\frac{M_B}{\sqrt{2}}x_1,b_1)+s(\frac{M_B}{\sqrt{2}}x_2,b_2)
+\frac{5}{3}\int_{1/b_1}^t\frac{d\mu}{\mu}\gamma_q(\mu)+2\int_{1/b_2}^t\frac{d\mu}{\mu}\gamma_q(\mu),\nonumber\\
S_{cd}(t)&=&s(\frac{M_B}{\sqrt{2}}x_1,b_2)+s(\frac{M_B}{\sqrt{2}}x_2,b_2)+s(\frac{M_B}{\sqrt{2}}x_3,b_3)+s(\frac{M_B}{\sqrt{2}}(1-x_3),b_3)
\nonumber\\&&+\frac{11}{3}\int_{1/b_2}^t\frac{d\mu}{\mu}\gamma_q(\mu)+2\int_{1/b_3}^t\frac{d\mu}{\mu}\gamma_q(\mu),\nonumber\\
S_{ef}(t)&=&s(\frac{M_B}{\sqrt{2}}x_2,b_2)+s(\frac{M_B}{\sqrt{2}}x_3,b_3)+s(\frac{M_B}{\sqrt{2}}(1-x_3),b_3)\nonumber\\&&
+2\int_{1/b_2}^t\frac{d\mu}{\mu}\gamma_q(\mu)+2\int_{1/b_3}^t\frac{d\mu}{\mu}\gamma_q(\mu),\nonumber\\
S_{gh}(t)&=&s(\frac{M_B}{\sqrt{2}}x_1,b_1)+s(\frac{M_B}{\sqrt{2}}x_2,b_2)+
s(\frac{M_B}{\sqrt{2}}x_3,b_2)+s(\frac{M_B}{\sqrt{2}}(1-x_3),b_2),\nonumber\\&&
+\frac{5}{3}\int_{1/b_1}^t\frac{d\mu}{\mu}\gamma_q(\mu)+4\int_{1/b_2}^t\frac{d\mu}{\mu}\gamma_q(\mu),
\end{eqnarray}
where the functions $s(Q,b)$ are defined in Appendix A of \cite{epjc45711}. $\gamma_q=-\alpha_s/\pi$ is the anomalous dimension of the quark.

\section{Light-Cone Distribution Amplitudes }\label{sec:c}
Here,  we specify the light-cone distribution amplitudes (LCDAs) for pseudoscalar and vector
mesons. The expressions of twist-2 LCDAs are \cite{prd81014022}
\begin{eqnarray}
\phi_P^A(x)&=&\frac{f_P}{\sqrt{6}}3x(1-x)[1+a_1^PC_1^{3/2}(t)+a_2^PC_2^{3/2}(t)+a_4^PC_4^{3/2}(t)],\nonumber\\
\phi_V(x)&=&\frac{f_V}{\sqrt{6}}3x(1-x)[1+a_{1V}^{\parallel}C_1^{3/2}(t)+a_{2V}^{\parallel}C_2^{3/2}(t)],\nonumber\\
\phi_V^T(x)&=&\frac{f^T_V}{\sqrt{6}}3x(1-x)[1+a_{1V}^{\perp}C_1^{3/2}(t)+a_{2V}^{\perp}C_2^{3/2}(t)],
\end{eqnarray}
and those of twist-3 ones are
\begin{eqnarray}
\phi_P^P(x)&=&\frac{f_P}{2\sqrt{6}}[1+(30\eta_3-\frac{5}{2}\rho_P^2)C_2^{1/2}(t)-(\eta_3\omega_3+\frac{9}{20}\rho_P^2(1+6a_2^P))C_4^{1/2}(t)],\nonumber\\
\phi_P^t(x)&=&\frac{f_P}{2\sqrt{6}}[1+6(5\eta_3-\frac{1}{2}\eta_3\omega_3-\frac{7}{20}\rho_P^2-\frac{3}{5}\rho_P^2a_2^P)(1-10x+10x^2)],\nonumber\\
\phi_V^t(x)&=&\frac{3f^T_V}{2\sqrt{6}}t^2, \phi_V^s(x)=-\frac{3f^T_V}{2\sqrt{6}}t, \phi_V^{\nu}(x)=\frac{3f^T_V}{8\sqrt{6}}(1+t^2),
 \phi_V^a(x)=-\frac{3f^T_V}{4\sqrt{6}}t,
\end{eqnarray}
where $t=2x-1$, $f_V$ and $f_V^T$ are the decay constants of the vector meson with longitudinal and transverse polarization,
respectively. For all pseudoscalar mesons, we choose $\eta_3=0.015$ and $\omega_3=-3$ \cite{zpc48239}.
The mass ratio $\rho_{\pi(K)}=m_{\pi(K)}/m_0^{\pi(K)}$ and $\rho_{\eta_{q(s)}}=2m_{q(s)}/m_{qq(ss)}$, and
the Gegenbauer polynomials $C_n^{\nu}$(t) read
\begin{eqnarray}
C_2^{1/2}(t)&=&\frac{1}{2}(3t^2-1),\quad C_4^{1/2}(t)=\frac{1}{8}(3-30t^2+35t^4),\quad  C_1^{3/2}(t)=3t,
\nonumber\\ C_2^{3/2}(t)&=&\frac{3}{2}(5t^2-1),\quad C_4^{3/2}(t)=\frac{15}{8}(1-14t^2+21t^4).
\end{eqnarray}

\end{appendix}


\begin{thebibliography}{99}

\bibitem{iiba}
N. Brambilla et al., (Quarkonium Working Group), CERN-2005-005, hep-ph/0412158.
\bibitem{prd391342}
Dong-Sheng Du and Z. Wang, Phys. Rev. D \textbf{39}, 1342 (1989).
\bibitem{09095028}
Ho-Meoyng Choi and Chueng-Ryong Ji, Phys. Rev D \textbf{80}, 114003.
\bibitem{prd80054016}
Ho-Meoyng Choi and Chueng-Ryong Ji,
 Phys. Rev D \textbf{80}, 054016 (2009).

\bibitem{prd73054024}
Jia-Fu Liu  and Kuang-Ta Chao
Phys. Rev. D \textbf{56}, 4133 (1997);
M. A. Ivanov, J. G. K¡§orner and P. Santorelli, Phys. Rev. D
\textbf{73}, 054024 (2006).
\bibitem{prd77114004}
Junfeng Sun, Yueling Yang,Wenjie Du, and Huilan Ma, Phys. Rev. D
\textbf{77}, 114004 (2008).
\bibitem{prd81014022}
Xin Liu, Zhen-Jun Xiao, and Cai-Dian L\"{u}, Phys. Rev. D
\textbf{81}, 014022 (2010).
\bibitem{epjc45711}
Jian-Feng Cheng, Dong-Sheng Du, and Cai-Dian L\"{u} Eur. Phy.
J. C. \textbf{45}, 711 (2006).
\bibitem{epjc60107}
Junfeng Sun, Dong-sheng Du, Yueling Yang, Eur. Phy. J. C. \textbf{60},
107 (2009).
\bibitem{epjc63435}
Jun Zhang, Xian-Qiao Yu, Eur. Phy. J. C. \textbf{63}, 435 (2009).
\bibitem{chenghaiyang}
Hai-Yang Cheng and Chun-Khiang Chua, Phys. Rev. D \textbf{80},  114008 (2009).
\bibitem{epjc5705}
Dong-Sheng Du and Zheng-Tao Wei,  Eur. Phy. J. C. \textbf{5}, 705  (1998).
\bibitem{prd84074033}
Zhen-Jun Xiao and Xin Liu, Phys.Rev. D \textbf{84}, 074033 (2011).

\bibitem{prl744388}
Y. Y. Keum, H. n. Li and A. I. Sanda, Phys. Lett. B \textbf{504}, 6 (2001)
[arXiv:hep-ph/0004004]; C.-D. L\"u, K. Ukai, M.Z. Yang,
hep-ph/0004213, Phys. Rev. D {\bf 63}, 074009 (2001);
 C.-D. L\"u,  M.Z. Yang,   Eur. Phys.
J. C. {\bf 23}, 275 (2002).

\bibitem{0505020}
 B.H. Hong and C.D. Lu, Sci. China G \textbf{49}, 357-366 (2006).
\bibitem{08031073}
 Run-Hui Li, Cai-Dian L\"{u}, and Hao Zou, Phys. Rev. D \textbf{78}, 014018 (2008);
 Run-Hui Li, Cai-Dian L\"{u}, A.I. Sanda and Xiao-Xia Wang, Phys. Rev. D \textbf{81}, 034006 (2010).
\bibitem{09081856}
Hao Zou, Run-Hui Li, Xiao-Xia Wang, and Cai-Dian  L\"{u}, J. Phys. G: Nucl. Part. Phys. \textbf{37}, 015002 (2010).
\bibitem{0512347}
Ying Li, Cai-Dian L\"{u} and Cong-Feng Qiao, Phys.Rev.D \textbf{73},094006 (2006);
Ying Li and Cai-Dian L\"{u}, J.Phys.G \textbf{29}, 2115 (2003);
Ying Li, Cai-Dian L\"{u} and Zhen-Jun Xiao, J.Phys. G \textbf{31},  273 (2005).
\bibitem{0305335}
Yong-Yeon Keum, T. Kurimoto, Hsiang-nan Li, Cai-Dian L\"{u} and A.I. Sanda,
Phys. Rev. D \textbf{69},  094018 (2004).
\bibitem{0112127}
Cai-Dian L\"{u}, Eur. Phys. J. C \textbf{24}, 121 (2002).
\bibitem{prd67054028}
 T. Kurimoto, H. n. Li and A. I. Sanda,
Phys. Rev. D \textbf{67}, 054028 (2003).




\bibitem{rmp681125}
G. Buchalla, A.J. Buras, M.E. Lautenbacher, Rev. Mod. Phys.
\textbf{68}, 1125 (1996).
\bibitem{bbns}
M. Beneke, G. Buchalla, M. Neubert, and C.T. Sachrajda, Phys. Rev. Lett. \textbf{83}, 1914 (1999);
Nucl. Phys. B \textbf{591}, 313 (2000).
\bibitem{scet}
C.W.Bauer, S. Fleming, D. Pirjol and I. W. Stewart, Phys. Rev. D \textbf{63}, 114020 (2001) [arXiv:
hep-ph/0011336]; C.W.Bauer, D. Pirjol and I. W. Stewart, Phys. Rev. Lett. \textbf{87}, 201806 (2001)
[arXiv: hep-ph/0107002].
\bibitem{npb193381}
J. C. Collins and D. E. Soper, Nucl. Phys. B \textbf{193},
381 (1981); J. Botts and G. Sterman, Nucl. Phys. B  \textbf{325}, 62
(1989).






\bibitem{zpc48239}
V.M. Braun and I.E. Filyanov , Z. Phys. C \textbf{48}, 239 (1990);
P. Ball, V.M. Braun, Y. Koike, and K. Tanaka, Nucl. Phys. B
\textbf{529}, 323 (1998); P. Ball, J. High Energy Phys. \textbf{01},
010 (1999).
\bibitem{epjc28515}
 Cai-Dian  L\"{u}, M.-Z. Yang, Eur. Phys. J. C \textbf{28}, 515 (2003).
\bibitem{hqet}
 A. V. Manohar and M. B. Wise, Camb. Monogr. Part. Phys. Nucl. Phys. Cosmol. \textbf{10}, 1 (2000).
\bibitem{prd074004}
B. Melic, B. Nizic, and K. Passek, Phys. Rev. D \textbf{60}, 074004
(1999).


\bibitem{prd58114006}
Th. Feldmann, P. Kroll, and B. Stech, Phys. Rev. D \textbf{58},
114006 (1998).
\bibitem{prd48339}
R. Escribano and J.M. Frere, J. High Energy Phys. \textbf{06}
(2005) 029; J. Schechter, A. Subbaraman, and H. Weigel,
Phys. Rev. D \textbf{48}, 339 (1993).

\bibitem{npp37}
 Particle Data Group, J. Phys. G: Nucl. Part. Phys. \textbf{37}, 075021 (2010).
\bibitem{prd71014015}
P.Ball and R. Zwicky,  Phys. Rev. D \textbf{71}, 014015
(2005); P. Ball and R. Zwicky, J. High Energy Phys.\textbf{ 04} (2006);
P. Ball and G.W. Jones, J. High Energy Phys. \textbf{03}, 069 (2007).

\bibitem{wang} W. Wang, Y.-L. Shen, C.-D. Lu, Phys. Rev.  D79 (2009) 054012,
e-Print: arXiv: 0811.3748 [hep-ph]
\bibitem{prd60094014}
A. Ali, G. Kramer and C.-D, L\"u,  Phys. Rev. D {\bf 58},
094009 (1998);
 Yaw-Hwang Chen, Hai-Yang Cheng, B. Tseng, and Kwei-Chou Yang, Phys. Rev. D \textbf{60},  094014 (1999).
\bibitem{plb525240}
 Sebastien Descotes-Genon, Jibo He, Emi Kou and  Patrick Robbe,
 Phys. Rev. D \textbf{80} 114031 (2009).
\bibitem{rhorho}
 Ying Li and Cai-Dian L\"{u},        Phys. Rev. D \textbf{73}, 014024 (2006).
\bibitem{prd66054013}
 C.-H. Chen, Y.-Y. Keum and H.-N. Li, Phys. Rev. D \textbf{66}, 054013 (2002).
\bibitem{0508080}
Han-Wen Huang, Cai-Dian L\"{u}, Toshiyuki Morii, Yue-Long Shen, Ge-Liang Songe and Jin Zhu,
Phys. Rev. D \textbf{73}, 014011 (2006).

\end{thebibliography}
\end{document}